\documentclass[a4paper,11pt]{article}
\usepackage{jheppub} 
\usepackage{lineno}
\usepackage{array}
\usepackage{multirow}
\usepackage{arydshln}

\usepackage{tikz}
\usepackage{tikz-feynman}

\newcommand{\mt}[1]{\textrm{\tiny #1}}
\newcommand{\non}{\nonumber \\}

\usepackage{mathtools}
\usepackage{epstopdf}

\title{Tidal Response and Thermodynamics of Black Holes}

\author{Itamar Cohen,}
\author{Dina Meylakh,}
\author{Michael Smolkin}
\author{and Israel Warszawiak}

\affiliation{The Racah Institute of Physics, The Hebrew University of Jerusalem, \\ Jerusalem 91904, Israel \\}

\emailAdd{itamar.cohen14@mail.huji.ac.il}
\emailAdd{dina.meylakh@mail.huji.ac.il}
\emailAdd{michael.smolkin@mail.huji.ac.il}
\emailAdd{Israel.Warszawiak@mail.huji.ac.il}

\abstract{In this work, we revisit black hole Love numbers from two complementary perspectives. First, we develop a manifestly gauge-invariant framework that directly integrates out the short-distance degrees of freedom of a static black hole in arbitrary spacetime dimensions. This approach yields the effective point-particle action and its associated Love numbers without relying on the standard matching procedure or on the Regge–Wheeler equation and its associated master field. Second, we investigate the role of Love numbers in black hole thermodynamics by analyzing a Schwarzschild black hole subjected to various types of external perturbations. We show that Love numbers govern the induced polarization of the black hole and control the leading corrections to its thermodynamic properties, thereby clarifying their physical significance in black hole thermodynamics.}

\begin{document}
\maketitle
\flushbottom

\section{Introduction}
\label{sec:intro}

Love numbers \cite{Love1909} quantify tidal deformations and describe how a celestial object responds to an external gravitational field. In recent years, they have attracted significant attention across astrophysical and theoretical contexts, driven in part by the success of gravitational-wave observations \cite{LIGOScientific:2016aoc,LIGOScientific:2025jau}. In the case of black holes, Love numbers are of particular theoretical interest, as they probe fundamental aspects of black hole physics and General Relativity (GR).

By construction, Love numbers allow one to replace an extended object with a point particle endowed with a set of coefficients that encode its response to external tidal fields. In field-theoretic terms, this procedure corresponds to an Effective Field Theory (EFT) description, in which short-distance physics associated with the size of the object is integrated out and replaced by effective gauge-invariant couplings to long-wavelength degrees of freedom. In classical GR, this framework was introduced in \cite{Goldberger:2004jt,Goldberger:2007hy} and, in the present context, is known as the effective point-particle description of black holes, in which finite-size effects are systematically captured by higher-dimensional operators in the worldline action. More broadly, the EFT approach has proven to be a powerful and flexible tool for GR calculations involving compact objects, enabling a clear separation of scales and an efficient organization of tidal effects \cite{Porto:2005ac,Kol:2007rx,Kol:2007bc,Rothstein:2014sra,Porto:2016pyg,Levi:2018nxp,Goldberger:2022rqf}.

Love numbers also play an important role in gravitational-wave phenomenology, as tidal effects modify both the orbital dynamics and the emitted signal of inspiraling compact binaries; see, for example, \cite{Flanagan:2007ix,Cardoso:2017cfl,Cardoso:2019vof,Steinhoff:2021dsn}. The first black hole Love numbers were computed for the Schwarzschild solution under static tidal perturbations \cite{Damour:2009vw,Binnington:2009bb,Damour:2009b,Kol:2011vg}. These results have since been extended in several directions, including different types of perturbations and more general black hole solutions. Early developments include the computation of magnetic Love numbers for the four-dimensional Schwarzschild black hole \cite{Damour:2009vw,Binnington:2009bb}, which describe the response to external gravito-magnetic fields. More recently, these results have been generalized to static black holes in arbitrary spacetime dimensions, described by the Tangherlini solution \cite{Tangherlini:1963bw}, with corresponding magnetic Love numbers obtained in \cite{Hui:2020xxx,Hadad:2024lsf}.

Further generalizations include non-gravitational response coefficients, such as scalar and electromagnetic Love numbers \cite{Kol:2011vg,Hui:2020xxx}. Additional developments encompass finite-frequency effects \cite{Correia:2024jgr,Caron-Huot:2025tlq,Combaluzier--Szteinsznaider:2025eoc,Ivanov:2026icp,Bhattacharyya:2026itm}, more general black hole backgrounds \cite{Brustein:2021bnw,Barbosa:2025uau,Charalambous:2025ekl,Bhattacharyya:2025slf,Barbosa:2026qcv,Wang:2026qst} – most notably rotating (Kerr) black holes \cite{Poisson:2014gka,Pani:2015hfa,Pani:2015nua,LeTiec:2020spy,Chia:2020yla,Glazer:2024eyi} – as well as nonlinear tidal response coefficients \cite{Gurlebeck:2015xpa,Poisson:2020vap,DeLuca:2023mio,Riva:2023rcm,Iteanu:2024dvx}.\footnote{For a comprehensive overview spanning both recent and past publications, see \cite{Rodriguez:2026iot,Chakraborty:2026qru}.}

A particularly striking result is that black hole Love numbers vanish identically in four spacetime dimensions \cite{Damour:2009vw,Binnington:2009bb,Damour:2009b,Kol:2011vg}. This observation has motivated proposals for an underlying symmetry – often referred to as Love symmetry – responsible for this behavior \cite{Porto:2016zng,Charalambous:2021kcz,Hui:2021vcv,Hui:2022vbh,Charalambous:2022rre,Rai:2024lho,Combaluzier-Szteinsznaider:2024sgb,Kehagias:2024rtz,Gounis:2024hcm}. More recently, this symmetry has been further extended in \cite{Parra-Martinez:2025bcu}.

The EFT construction for compact objects in classical GR begins by identifying the relevant long-distance degrees of freedom and imposing the symmetries of the problem, namely general covariance and reparametrization invariance. One then writes down the most general effective action consistent with these symmetries, organized as an expansion in fields and their derivatives. Each operator is suppressed by a short-distance scale–the EFT cutoff–associated with the physics that has been integrated out. Lower-derivative operators dominate at leading order, while higher-derivative terms encode subleading finite-size effects \cite{Goldberger:2004jt,Goldberger:2007hy,Porto:2005ac,Kol:2007rx,Kol:2007bc,Rothstein:2014sra,Porto:2016pyg,Levi:2018nxp,Goldberger:2022rqf}. The corresponding Wilson coefficients are gauge-invariant and can be identified with Love numbers.

In the standard approach, Love numbers are determined by matching EFT calculations to results obtained in the full theory of GR. In this work, we instead show how to explicitly integrate out the short-distance degrees of freedom, thereby directly deriving the effective action terms associated with scalar, electromagnetic, and gravito-magnetic Love numbers for black holes in arbitrary spacetime dimensions. This method bypasses the traditional matching procedure and does not rely on the Ishibashi–Kodama master field. It is manifestly gauge invariant, as it avoids introducing propagators for long-distance fields and thus eliminates the need for gauge fixing. The resulting Love numbers are in full agreement with known results, and the method can be readily extended to other types of perturbations.

The second focus of this paper is the connection between Love numbers and black hole thermodynamics \cite{Bekenstein:1973ur,Bardeen:1973gs}. The thermodynamic description of a black hole depends strongly on the asymptotic structure of spacetime \cite{Gibbons:1977,FrolovZelnikov2011}. Assumptions about asymptotics determine how quantities such as energy, temperature, and entropy are defined, as well as the appropriate thermodynamic ensemble. While one typically assumes asymptotically flat, Anti–de Sitter, or de Sitter boundary conditions, such assumptions tend to obscure the role of Love numbers, which become physically relevant only in the presence of external fields.

The two parts of this work are based on the same underlying principle: the response of a black hole to an external field can be extracted from the on-shell action. In the first part, we use this principle to derive the worldline EFT directly from the underlying theory. The derivation is top-down and therefore does not require gauge fixing or a diagrammatic expansion: the EFT coefficients are identified by evaluating the on-shell action in the presence of external fields. In the second part, we apply the same on-shell-action viewpoint to the thermodynamics of black holes in external environments. Here, gauge fixing and Feynman diagrams are employed as calculational tools for evaluating the relevant thermodynamic quantities and organizing the perturbative expansion. Their use does not represent a change in the underlying physical framework, but rather reflects the fact that the thermodynamic calculation is carried out within a perturbative description of the gravitational and matter fields. Thus, the two approaches should be viewed as complementary: the first establishes the general EFT interpretation of the response, while the second provides a thermodynamic characterization of the same response.

More concretely, the response captured by the EFT and the polarization variable appearing in the thermodynamic description encode the same physical response to the external field. In the EFT language, the response is characterized by the Wilson coefficient of a worldline operator quadratic in the external field, while variation of the corresponding on-shell effective action with respect to the external source defines the induced multipole moment. In the thermodynamic description, the polarization is obtained as the variable conjugate to the external multipole in the appropriate thermodynamic potential. The two descriptions therefore provide complementary ways of organizing the same response: the EFT identifies its local worldline operator and Wilson coefficient, while the thermodynamic formulation relates the corresponding response to variations of the black-hole free energy.

In this work, we consider a Schwarzschild black hole in an asymptotically flat spacetime and study how its thermodynamic properties are modified by external scalar, electric, and weakly curved gravitational fields. This setup is analogous to linear response in ordinary materials subjected to external electromagnetic perturbations. In such systems, external fields couple to microscopic degrees of freedom, inducing polarization, reducing disorder, and modifying thermodynamic potentials. Similarly, external fields interacting with a black hole induce polarization effects encoded by Love numbers.

Since Love numbers characterize the linear response to external perturbations, they directly determine the induced polarization and the corresponding corrections to thermodynamic quantities. In what follows, we derive explicit relations between Love numbers and thermodynamic properties of a Schwarzschild black hole in the presence of scalar and electric, as well as gravito-electric and gravito-magnetic perturbations.

This paper is organized as follows. In section \ref{sec:preliminary}, we review a useful metric decomposition, introduce the effective action, and summarize relevant aspects of black hole thermodynamics. In section \ref{sec:smarr}, we analyze the first law in the presence of external fields. Section \ref{sec:example} presents a simple example from classical electrodynamics to illustrate the direct integration of short-distance degrees of freedom. Sections \ref{sec:scalar_bckgrnd}–\ref{sec:gravito-electric} apply this method to various types of perturbations of a Schwarzschild black hole. We conclude with a summary and discussion, and provide additional technical details in the appendices.

\section{Preliminary results}
\label{sec:preliminary}

This section presents background material that will be helpful throughout the paper. We introduce the notation and definitions used in what follows and recall a number of standard results that will be referenced frequently. The aim is not to provide exhaustive derivations, but to fix terminology, establish conventions, and make the presentation self-contained. Readers familiar with the subject may safely skip the introductory part of this section, as well as subsections~\ref{sec:BH-EFT} and~\ref{sec:GRstresstensor}, returning to them as needed for reference.

Following \cite{Kol:2007rx}, we implement the following parametrization of the metric
\begin{equation}
ds^{2}=g_{\mu\nu}dx^{\mu}dx^{\nu}=e^{2\phi}\left(dt-2 A_{I}dx^{I}\right)^{2}-e^{-\frac{2\phi}{d-3}}\gamma_{IJ}dx^{I}dx^{J} ~,
\label{KK}
\end{equation}
where capital Latin indices run over the $(d-1)$-dimensional space, with $I, J = 1, 2, \ldots, d-1$. This change of variables is particularly well suited to the time-independent metrics studied in this work, in which case the Einstein–Hilbert action reduces to\footnote{Throughout this paper, we use units in which $G=c=1$.}
\begin{equation}
S_\mt{EH}\left(\phi,A,\gamma\right)=-\frac{1}{16\pi}\int dt\int d^{d-1}x\,\sqrt{\gamma}\,\big(- \mathcal{R}\left[\gamma\right]+\frac{d-2}{d-3}\partial_{I}\phi\partial_{J}\phi\gamma^{IJ}- e^{2\frac{d-2}{d-3}\phi}F_{IJ}F_{KL}\gamma^{IK}\gamma^{JL}\big) ~,
\label{eq:EH_action}
\end{equation}
where $F_{IJ}=\partial_{I}A_{J}-\partial_{J}A_{I}$ and $\mathcal{R}\left[\gamma\right]$ is the Ricci scalar of the spatial metric $\gamma_{IJ}$. Note that the integral over time factors out and may be omitted.

In particular, the equations of motion for time-independent geometries are given by
\begin{align}
 &\mathcal{R}_{IJ} - {\gamma_{IJ} \over 2} \mathcal{R} =  {d-2\over d-3} T^\phi_{IJ} -2 \, e^{2\frac{d-2}{d-3}\phi} T^\mt{EM}_{IJ}~,
 \non
 &   {1\over \sqrt{\gamma}}\partial_I \Big( \sqrt{\gamma} \, \gamma^{IJ} \,\partial_J\phi\Big) + e^{2\frac{d-2}{d-3}\phi} F^2 =0~,
\label{eq:EOM_NRG}
 \\
 & \nabla_I \Big(e^{2\frac{d-2}{d-3}\phi} F^{IJ}\Big) =0 ~,
 \nonumber
\end{align}
where $\nabla_I$ denotes the covariant derivative compatible with $\gamma_{IJ}$, and the stress-energy tensors are 
\begin{align}
    T_{IJ}^{\phi}&=\partial_{I}\phi\partial_{J}\phi - \frac{\gamma_{IJ}}{2}\left(\partial\phi\right)^{2} ~,
\non
    T_{IJ}^\mt{EM}&=F_{IA}F_{JB}\gamma^{AB} 
    - \gamma_{IJ} \Big( {1\over 4} F^2 \Big) ~.
\label{eq:EOM_NRG_Tensors_phi}
\end{align}

For static geometries, $A_I$ vanishes. An example of such a geometry is provided by the Schwarzschild black hole metric in a general number of spacetime dimensions, commonly known as the Tangherlini solution \cite{Tangherlini:1963bw},
\begin{equation}
ds^{2}=f\left(r\right)dt^{2}-\frac{1}{f\left(r\right)}dr^{2}-r^{2}d\Omega_{d-2}^{2} \quad , \quad 
f\left(r\right)=1-\left(\frac{r_{s}}{r}\right)^{d-3} ~,
\label{eq:Schw}
\end{equation}
where $r_{s}$ is the Schwarzschild radius. We denote the fields that describe the Schwarzschild black hole by $\phi_S$ and $\gamma^S_{IJ}$. Using \eqref{KK}, one can read off their explicit expressions 
\begin{equation}
 \phi_S= {1\over 2} \log f(r) ~, \quad \gamma^S_{rr}=f^{-{d-4\over d-3}}~, \quad \gamma^S_{ij} = f^{1\over d-3} r^2 \, \Omega_{ij}~.
 \label{Sch metric}
\end{equation}
Here and in the following sections, lower-case Latin indices run over the $(d-2)$-dimensional sphere, and $\Omega_{ij}$ represents the metric on a unit sphere.

\subsection{Black hole effective action}
\label{sec:BH-EFT}

Consider a Schwarzschild black hole \eqref{eq:Schw} probed at distances much larger than $r_s$. In this regime, the black hole can be approximated as a point particle coupled to a long-wavelength gravitational field. However, this approximation neglects finite-size effects and the internal structure of the black hole. To account for these corrections systematically, we adopt the EFT approach introduced in \cite{Goldberger:2004jt}.

Within the EFT framework, the full general-relativistic dynamics are captured by an effective action, with finite-size effects encoded in an infinite tower of generally covariant operators localized on the worldline of the point-like black hole. The first few purely gravitational terms, containing up to four derivatives, in the resulting effective action are given by
\begin{align}
\label{eq:PP_action}
S_\text{p.p.}=-m\int d\tau+\frac{C_{E}}{4}\int d\tau E_{\mu\nu}E^{\mu\nu}+\frac{C_{B}}{4}\int d\tau B_{\alpha_{1}...\alpha_{d-2}}B^{\alpha_{1}...\alpha_{d-2}}+...\,\,,
\end{align}
where $\tau$ is the proper time of the point-like black hole moving along the worldline trajectory $x^\mu\left(\tau\right)$ and\footnote{A completely antisymmetric tensor $\varepsilon_{\mu_1\cdots\mu_d}$ is defined as $$\varepsilon_{\mu_1\cdots\mu_d}=\sqrt{|g|} \epsilon_{\mu_1\cdots\mu_d} ~,$$ where $\epsilon_{\mu_1\cdots\mu_d}$ is the Levi–Civita symbol – a tensor density of weight 1 – normalized by $\epsilon_{01\cdots d-1}=+1$.} 
\begin{gather}
\label{eq:gravito_EM_components_E}
E_{\mu\nu}=\mathcal{R}_{\mu\alpha\nu\beta}\dot{x}^{\alpha}\dot{x}^{\beta},\,\,\,\dot{x}^{\alpha}=\frac{dx^{\alpha}}{d\tau} ~, \\
\label{eq:gravito_EM_components_B}
B_{\alpha_{1}...\alpha_{d-2}}=\frac{1}{\left(d-2\right)!}\dot{x}^{\sigma}\dot{x}^{\rho}\varepsilon_{\sigma\alpha_{1}...\alpha_{d-3}\mu\nu}\mathcal{R}_{\,\,\,\rho\alpha_{d-2}}^{\mu\nu} ~,
\end{gather}
are the gravito-electric and gravito-magnetic components of the Riemann tensor. The coefficients \(C_{E}\) and \(C_{B}\) characterize the electric-type and magnetic-type quadrupole susceptibilities of a spinless black hole, respectively. 

If the black hole is stationary and the gravitational fields are weak, the effective action simplifies. Using the parameterization \eqref{KK} and retaining quadratic terms in weak fields - which are sufficient for a linear response analysis - we obtain\footnote{For the remainder of this section, Latin indices are raised and lowered using the Kronecker delta $\delta_{IJ}$.}
\begin{align}
\label{eq:PP_action_weak_fields}
S_\mt{p.p.}=-m\int dt \Big(\phi-\frac{\phi^{2}}{2}\Big)+\sum_{l=2}^{\infty}\frac{C_{l}^{E}}{2l!}\int dt \, \partial_{I_{1}}...\partial_{I_{l}}\phi ~ \partial^{I_{1}}...\partial^{I_{l}}\phi \notag \\ +\sum_{l=2}^{\infty}\frac{C_{l}^{B}}{2l!} \int dt \, \partial_{I_{1}}...\partial_{I_{l-1}}F_{MN}~\partial^{I_{1}}...\partial^{I_{l-1}}F^{MN},
\end{align}
where the coefficients \(C_{l}^{E}\) and \(C_{l}^{B}\) are referred to as Love numbers for the electric-type and magnetic-type susceptibilities of the black hole, with $l$ representing the multipole order. As shown in \cite{Kol:2011vg,Hui:2020xxx,Hadad:2024lsf}, these Love numbers are given by\footnote{The coefficients $C_E, C_B$ in \eqref{eq:PP_action} are directly related to the $l=2$ coefficients $C_l^E$ and $C_l^B$ $$C_E=C^E_{l=2}\, , \quad C_B=(-1)^d{(d-2)(d-2)!\over 2} \, C_{l=2}^B ~.$$ while higher-$l$ coefficients are associated with operators contained in the ellipsis of \eqref{eq:PP_action}, constructed by squaring expressions involving $l-2$ derivatives of the gravito-electric and gravito-magnetic components \eqref{eq:gravito_EM_components_E} and \eqref{eq:gravito_EM_components_B}, respectively.}
\begin{align}
C_{l}^{E}&=\frac{\Omega_{d-3+2l} }{(2\pi)^l }~
\frac{(d-2) (l+d-2)}{(d-3)(l-1)} ~
\frac{16^{-\,\hat l-1}\,B\big(\,\hat l \,,\, \hat l+2\big)}{B\big( \, \hat l+\frac{1}{2} \, , \, \hat l+\frac{3}{2}\big)} ~
\tan \big(\pi \hat l\, \big) \, r_{s}^{d-3+2l} ~, \quad \hat l={l\over d-3} ~,
\nonumber\\
 C_l^B &= -{\Omega_{d-3+2l}\over 8 (2\pi)^l}~
 {1+\hat{l}_+\over 1+\hat{l}_+ + \hat{l}_-} ~
 {B(\hat{l}_+,\hat{l}_-) \over B(-\hat{l}_+\,, \, -\hat{l}_-)} ~
 r_{s}^{d-3+2l} ~, \quad 
 \hat{l}_\pm={l\pm 1\over d-3} ~.
 \label{eq:LoveNum}
\end{align}
where $\Omega_d$ represents the area of a unit sphere in $d$-dimensional space, and $B(x,y)$ is the beta function,
\begin{equation}
\Omega_{d}={2\pi^{d\over 2} \over \Gamma\big({d\over 2}\big)}  ~, \quad B(x,y) ={ \Gamma( x) \, \Gamma( y) \over \Gamma(x+y) } ~.
\label{eq:OmegaDef}
\end{equation}

To compute classical amplitudes of the metric components within the EFT formalism, a gauge choice is necessary. We adopt a linearized version of the harmonic gauge, $g^{\mu\nu}\Gamma_{\mu\nu}^{\alpha} = 0$. In this gauge, the action is supplemented by a gauge-fixing term which, when expressed in terms of the variables \eqref{KK}, takes the simple form
\begin{align}
\label{eq:GF_action}
S_{\textrm{GF}}=\frac{1}{32\pi}\int dt\int d^{d-1}x\left[4\left(\partial^{J}A_{J}\right)^{2}-\left(\frac{1}{2}\partial_{I}\sigma_{\,J}^{J}-\partial_{J}\sigma_{I}^{\,J}\right)^{2}\right], \quad \sigma_{IJ}=\gamma_{IJ}-\delta_{IJ}~,|\sigma_{IJ}|\ll 1 ~.
\end{align}
The full EFT action is given by
\begin{align}
\label{eq:full_EFT_action}
S_{\textrm{EFT}}=S_{\textrm{EH}}+S_{\textrm{GF}}+S_{\textrm{p.p.}}.
\end{align}
The propagators of the various fields in this gauge take the form\footnote{To compute the $\sigma_{IJ}$ propagator, one needs to find the inverse of the quadratic term involving $\sigma_{IJ}$ in the Einstein-Hilbert action \eqref{eq:EH_action}, which is supplemented with the gauge fixing term \eqref{eq:GF_action}. This involves inverting the tensor structure of the form $\delta^{IK}\delta^{JL}+\delta^{IL}\delta^{JK}-\delta^{IJ}\delta^{KL}$ in the vector space of symmetric rank 2 tensors, resulting in $\frac{1}{4}\big(\delta_{IK}\delta_{JL}+\delta_{IL}\delta_{JK}-\frac{2}{\left(d-3\right)}\delta_{IJ}\delta_{KL}\big)$.}
\begin{align}
&\left\langle \phi\left(x\right)\phi\left(0\right)\right\rangle =-\frac{4\left(d-3\right)}{\left(d-2\right)\Omega_{d-3}|x|^{d-3}}~,
\label{eq:propPhi}\\
&\left\langle A_I(x)A_J(0)\right\rangle = \frac{2}{\Omega_{d-3}} \, {\delta_{IJ} \over |x|^{d-3}} ~,
\label{eq:propA} \\
&\left\langle \sigma_{IJ}\left(x\right)\sigma_{KL}(0)\right\rangle =-\frac{8}{\Omega_{d-3}|x|^{d-3}}\left(\delta_{IK}\delta_{JL}+\delta_{IL}\delta_{JK}-\frac{2}{d-3}\delta_{IJ}\delta_{KL}\right)~.
\label{eq:propSigma}
\end{align}
Throughout this paper, the propagators of the scalar, vector, and tensor fields are represented in Feynman diagrams by solid, dashed, and wiggly lines, respectively.

As usual, the construction of the effective action presented above relies solely on the underlying symmetry principles –general covariance and reparametrization invariance – rather than on an explicit derivation from the full theory of general relativity. In general, obtaining the effective action directly from the microscopic dynamics is not feasible. Nevertheless, as we demonstrate in Sections~\ref{sec:scalar_bckgrnd}, \ref{sec:maxwell_bckgrnd} and \ref{sec:gravito-magnetic}, a direct derivation of \eqref{eq:PP_action_weak_fields} is in fact tractable.

An effective action constructed on symmetry grounds is therefore determined only up to a set of a priori unknown coefficients, such as \(C_{l}^{E}\) and \(C_{l}^{B}\) in \eqref{eq:PP_action_weak_fields}, commonly referred to as Wilson coefficients, which multiply the allowed operators. These coefficients encode the effects of short-distance physics that has been integrated out and cannot be fixed by symmetry considerations alone. Their values are determined through a matching procedure, in which physical observables or correlation functions computed in the effective theory are equated with those obtained in the underlying microscopic theory. This ensures that the effective action faithfully reproduces the long-wavelength behavior of the full theory.

The full effective action \eqref{eq:full_EFT_action} will be used extensively in the following sections to compute the asymptotic form of the metric, which plays a central role in the thermodynamic analysis of the black hole. As an illustration, in the next subsection we recall the precise relation between the internal energy of the gravitational system and the asymptotic behavior of the metric.

\subsection{Gravitational Pseudotensor}
\label{sec:GRstresstensor}

In this work, the metric $g_{\mu\nu}$ approaches the Minkowski metric $\eta_{\mu\nu}$ at large distances from the black hole. That is,
\begin{equation}
    g_{\mu\nu}=\eta_{\mu\nu}+\delta g_{\mu\nu} ~,
\end{equation}
where $\delta g_{\mu\nu}$ need not be small everywhere but becomes small asymptotically, i.e., $|\delta g_{\mu\nu}|\ll 1$ at large distances. In the absence of a weakly curved external background, $\delta g_{\mu\nu}$ vanishes at infinity; otherwise, it remains small far from the black hole. 

In this setting, the total energy and momentum of the system can be expressed in terms of the conserved total energy-momentum pseudotensor $\tau^{\mu\nu}=T^{\mu\nu}+t^{\mu\nu}$, where $T^{\mu\nu}$ is the matter energy-momentum tensor and $t^{\mu\nu}$ is the energy-momentum pseudotensor of the gravitational field itself \cite{landau1994classical,Weinberg:1972kfs}
\begin{equation}
    t_{\mu\nu}={-1\over 8\pi}\Big(\mathcal{R}_{\mu\nu} - {1\over 2}g_{\mu\nu} \mathcal{R} - 
    \mathcal{R}^{(1)}_{\mu\nu} + {1\over 2}\eta_{\mu\nu}\mathcal{R}^{(1)} \Big) ~,
\end{equation}
where\footnote{The indices of the pseudotensors $\tau^{\mu\nu}$ and $t^{\mu\nu}$ are raised and lowered using $\eta_{\mu\nu}$.}
\begin{equation}
    \mathcal{R}^{(1)}_{\mu\nu}= {1\over 2}
\big(\partial^\lambda\partial_{\mu}\delta g_{\lambda\nu}+\partial^\lambda\partial_{\nu}\delta g_{\lambda\mu} - \partial_\mu\partial_\nu \delta g^\lambda_\lambda - \partial^2\delta g_{\mu\nu}\big)~,
    \quad \mathcal{R}^{(1)}=\eta^{\mu\nu}\mathcal{R}^{(1)}_{\mu\nu} ~,
\end{equation}
is the part of the Ricci tensor linear in $\delta g_{\mu\nu}$. 

Unlike the matter energy-momentum tensor $T^{\mu\nu}$, which satisfies a covariant conservation law that describes the exchange of energy between matter and gravity, the total energy-momentum pseudotensor is conserved in the ordinary sense
\begin{equation}
    \partial_\mu \tau^{\mu\nu}=0~.
\end{equation}
Thus, the conserved total energy-momentum pseudovector is given by
\begin{equation}
    P^\mu=\int d^{d-1}x \, \tau^{0\mu}  ~.
\end{equation}
In particular, after some algebra, the energy $U=P^0$ takes the form \cite{Weinberg:1972kfs}
\begin{equation}
    U={1\over 16\pi}\int \Big( \partial_I \delta g_{J}^J -\partial_J \delta g_{I}^J \Big) n^I R^{d-2}d\Omega~.
\end{equation}
The integral is taken over a large sphere of radius $R$, where $n^I$ denotes the outward-pointing unit normal vector and $d\Omega$ is the differential solid angle. When expressed in terms of the fields in \eqref{KK}, this expression becomes more lengthy due to the nonlinear relation between $\delta g_{\mu\nu}$ and $\phi, A_I$ and $\gamma_{IJ}$
\begin{align}
    U&={1\over 16\pi}\int \Big( \partial_I\Big( 2{d-2\over d-3}\phi-\sigma_{J}^J + 4A_JA^J\Big)+\partial_J\big(\sigma_{I}^J - 4A_IA^J\big) \Big) n^I R^{d-2}d\Omega 
    \label{eq:U_with_NRG}\\
    &+{1\over 16\pi}\int \Big( - {4(d-2)\over(d-3)^2} \phi\partial_I\phi +{2\over d-3}\partial_I(\phi\sigma_J^J) - {2\over d-3} \partial_J(\phi\sigma^J_I)\Big)n^I R^{d-2}d\Omega + \ldots~,
    \nonumber
\end{align}
where $\sigma_{IJ} = \gamma_{IJ} - \delta_{IJ}$ while $|\sigma_{IJ}|\ll 1$, and the ellipsis encodes cubic and higher-order terms in the small fields. We retain only quadratic terms, as this is sufficient for our purposes when studying the thermodynamic aspects of perturbed black hole.

In the absence of external perturbations, the vector field $A_I$ vanishes identically due to the $t\to-t$ symmetry, and the mass $m$ in the black hole effective action \eqref{eq:PP_action_weak_fields} is the sole source for the remaining gravitational fields $\phi$ and $\sigma_{IJ}$. The corresponding worldline vertex is

\vspace{1cm}

\begin{tikzpicture}
    \begin{feynman} [scale=1.0, transform shape]
        \vertex (a);
        \vertex [below=1cm of a] (b);
        \vertex [below=1cm of b] (c);
        \vertex [right=of b] (d);
        
        \diagram* {
            (a) -- [double,  style={/tikz/double distance=2pt}] (b),
            (b) -- [double,  style={/tikz/double distance=2pt}] (c),
            (b) -- [ very thick] (d),
        };
        \node [right, yshift=0mm] at ([xshift=0.7cm]d) {=\quad$m\phi\left(0\right)$~.};
        \vertex [dot, minimum size=2pt] at (b) {};
    \end{feynman}
\end{tikzpicture}

\vspace{1cm}\noindent
Attaching the scalar propagator to it yields 
\begin{equation}
    \phi(x)=-\frac{4\left(d-3\right)}{\left(d-2\right)\Omega_{d-3}} \, {m\over |x|^{d-3}} + \ldots ~,
    \quad 
    \sigma_{IJ}(x)= 0 + \ldots ~,
\end{equation}
where the ellipsis represents corrections that decay as $1/r^{2(d-3)}$ or faster, and therefore do not contribute to $U$.\footnote{The diagrams contributing to $\sigma_{IJ}$ carry at least two mass insertions and therefore decay rapidly at infinity.} Substituting this expression into \eqref{eq:U_with_NRG} yields the well-known result for the energy of the $d$-dimensional Schwarzschild black hole
\begin{equation}
    \label{eq:USch}
    U_\text{Sch}={(d-2)\over 8\pi(d-3)}\int \, n^I\, \partial_I\phi  \, R^{d-2}d\Omega =  m ~.
\end{equation}

In the following sections, weak stationary perturbations of the Schwarzschild black hole are introduced, and $S_\mt{EFT}$ is used to perturbatively evaluate the asymptotic values of $\phi$, $A_I$, and $\sigma_{IJ}$. Substituting these expressions into \eqref{eq:U_with_NRG} yields the leading-order corrections to the unperturbed energy $U_\text{Sch}$ of the Schwarzschild black hole in the presence of external background fields,
\begin{equation}
    \label{eq:Upert}
    U=U_\text{Sch} + U^{(1)} + U^{(2)}+\ldots = U_\text{Sch}+\sum_{n=1}^\infty U^{(n)} ~,
\end{equation}
where $U^{(n)}$ denotes the $n$-th order correction to $U_\text{Sch}$.

\subsection{Exact gravitational action and free energy}
\label{sec:exact_gravitational_action}

The seminal work of Gibbons and Hawking \cite{Gibbons:1977} highlights the crucial role played by the full gravitational action in black hole thermodynamics. In this subsection, we briefly review their main results and explain how they lead to the point-particle effective action \eqref{eq:PP_action_weak_fields}.

The classical gravitational action consists of three contributions: the Einstein–Hilbert term, the Gibbons–Hawking–York (GHY) boundary term, and the matter action describing non-gravitational degrees of freedom,
\begin{align}
\label{eq:full_exact_action}
S_{\textrm{full}}=S_{\textrm{EH}}+S_{\textrm{GHY}}+S_\text{matter}=-\frac{1}{16\pi}\int_{V}\sqrt{\left|g\right|}\mathcal{R}+\frac{1}{8\pi}\int_{\partial V}\sqrt{\left|h\right|}\left[\mathcal{K}\right]+S_\text{matter} ~.
\end{align}
The inclusion of the GHY term is essential, as the Einstein–Hilbert action alone leads to an ill-defined variational principle due to the presence of second derivatives of the metric in the Ricci scalar $\mathcal{R}$. In the above expression, $V$ is the bulk of a $d$-dimensional spacetime, and the boundary $\partial V$ is taken to be the direct product of time and a spatial sphere of radius $R$, i.e., $\partial V=\mathbb{R}\times S^{d-2}$. The timelike hypersurface $\partial V$ is equipped with an outward-pointing spacelike unit normal vector $n^{\mu}$. The quantity $\mathcal{K}$ denotes the extrinsic curvature of \(\partial V\) embedded in the spacetime with metric \(g_{\mu\nu}\), while \(h_{\mu\nu}\) is the induced metric on \(\partial V\). The square brackets indicate subtraction of a reference geometry. For brevity, we will suppress this subtraction term in what follows. In addition, we focus on perturbations of the Schwarzschild black hole. Therefore, in the region of interest, $S_\text{matter}$ either vanishes or represents a weak perturbation of the gravitational action.

To obtain the effective action, one typically decomposes the metric into long- and short-wavelength components and then integrates out the latter. In the classical gravity regime considered here, integrating out the short-wavelength modes amounts to evaluating the full action $S_{\textrm{full}}$ on solutions of their classical equations of motion, with the long-wavelength components acting as fixed sources or boundary data. In practice, however, deriving an effective action $S_{\textrm{EFT}}$ in this manner is not always feasible, owing to the strong interactions of the short-wavelength modes. One therefore adopts the standard approach outlined in subsection~\ref{sec:BH-EFT}, which is based on symmetry arguments and supplemented by a matching procedure.

Nevertheless, the effective action \eqref{eq:PP_action_weak_fields} can be derived directly from the full action \eqref{eq:full_exact_action}. To do so, consider a Schwarzschild black hole embedded in an external, weakly curved stationary gravitational background whose Ricci curvature satisfies 
\begin{equation}
    \mathcal{R}\ll 1/R^2 \ll 1/r_s^2 ~.
\end{equation} 
From the EFT perspective, this background acts as a long-wavelength gravitational field that is weakly coupled to the short-wavelength modes associated with the Schwarzschild scale $r_s$. From the viewpoint of the full theory, the external gravitational field modifies the boundary conditions far from the black hole horizon. Since the background curvature is weak, these modified boundary conditions generate only perturbative corrections to the Schwarzschild metric $g_{\mu\nu}^{S}$. Accordingly, the spacetime metric admits the expansion
\begin{align}
\label{eq:gmunu_perturbation}
g_{\mu\nu}=g_{\mu\nu}^{S}+g_{\mu\nu}^{\left(1\right)}+g_{\mu\nu}^{\left(2\right)}+...=g_{\mu\nu}^{S}+\sum_{n=1}^{\infty}g_{\mu\nu}^{\left(n\right)}=g_{\mu\nu}^{S}+\delta g_{\mu\nu}
\end{align}
where \(g_{\mu\nu}^{\left(n\right)}\) denotes the $n$–th order correction to $g_{\mu\nu}^{S}$, induced by the weak external background and controlled by the expansion parameter $\mathcal{R} R^2\ll 1$, and the final equality defines the total perturbation \(\delta g_{\mu\nu}\). 

The weakly curved background may be sourced either by matter located outside the region of interest $V$, in which case the matter action in \eqref{eq:full_exact_action} vanishes, or by matter fields $\psi$ living within $V$. In the latter case,  \eqref{eq:gmunu_perturbation} is accompanied by a corresponding expansion of $\psi$\footnote{The field $\psi$ is purely perturbative and therefore vanishes at zeroth order.}
\begin{align}
\label{eq:psi_perturbation}
\delta\psi=\psi^{\left(1\right)}+\psi^{\left(2\right)}+...=\sum_{n=1}^{\infty} \psi^{\left(n\right)} ~,
\end{align}
and the metric perturbation \eqref{eq:gmunu_perturbation} begins at second order, $g_{\mu\nu}^{\left(1\right)}=0$,  since the matter energy–momentum tensor that sources the weakly curved background is quadratic in $\psi$. 

By construction, $\delta g_{\mu\nu}$ matches onto the weakly curved background at the boundary $r=R$ and satisfies $\delta g_{\mu\nu}\ll 1$ throughout the bulk. Solving the Einstein equations perturbatively for $\delta g_{\mu\nu}$ and substituting the resulting solution into $S_{\textrm{full}}$ yields a perturbative expansion of the on-shell action. This expansion should be identified with the effective action describing the black hole coupled to long-wavelength modes,
\begin{align}
\label{eq:action_perturbation_def}
S_{\textrm{EFT}}=S_{\textrm{full}}\Big|_\text{on-shell}=S_{\textrm{Sch}}+[\delta^{\left(1\right)}S_{\textrm{full}}]+\frac{1}{2}[\delta^{\left(2\right)}S_{\textrm{full}}]+...=S_{\textrm{Sch}}+\sum_{n=1}^{\infty}\frac{[\delta^{\left(n\right)}S_{\textrm{full}}]}{n!}~,
\end{align}
where $S_{\textrm{Sch}}$ denotes the value of $S_{\textrm{full}}$ evaluated on the unperturbed Schwarzschild black hole, square brackets indicate the necessary subtractions, and $\delta^{\left(n\right)}S_{\textrm{full}}$ represents the $n$-th order correction induced by the weakly curved background. In what follows, we present a number of explicit examples illustrating that \eqref{eq:action_perturbation_def} reproduces the black hole effective action. In Sections~\ref{sec:scalar_bckgrnd} and \ref{sec:maxwell_bckgrnd}, we recover the effective action corresponding to the scalar and electric susceptibilities of the Schwarzschild black hole, with Love numbers in agreement with those reported in the literature. In Section~\ref{sec:gravito-magnetic}, we reproduce the magnetic sector of \eqref{eq:PP_action_weak_fields}, with $A_I$ identified as the long-wavelength field associated with the weakly curved background and with the corresponding gravito-magnetic Love numbers matching those given in \eqref{eq:LoveNum}.

Each order in the expansion \eqref{eq:action_perturbation_def} is obtained by taking the $n$-th variation of the exact action \eqref{eq:full_exact_action}. To perform these variations, we make use of the results in Appendix~\ref{app:curvature_variations}, where the variations of the relevant geometric quantities are derived and summarized. The first variation is given by \footnote{Our convention for the energy-momentum tensor is $$T^{\mu\nu}=-{2\over\sqrt{|g|}} {\delta S_\mt{matter}\over\delta g_{\mu\nu}}~.$$}
\begin{align}
\delta S_{\textrm{full}}=\frac{1}{16\pi}\int_{V}\sqrt{\left|g\right|}\Big(\mathcal{R}^{\mu\nu}-\frac{g^{\mu\nu}}{2}\mathcal{R} - 8\pi \, T^{\mu\nu}\Big)\delta g_{\mu\nu}+\frac{1}{16\pi}\int_{\partial V}\sqrt{\left|h\right|}\left(h^{\mu\nu}\mathcal{K}-\mathcal{K}^{\mu\nu}\right)\delta g_{\mu\nu}
\nonumber\\
+\int_{V} \sqrt{\left|g\right|} \Big({\partial\mathcal{L}_\text{matter}\over\partial\psi}-\nabla_\mu {\partial\mathcal{L}_\text{matter}\over\partial\nabla_\mu\psi}\Big)\delta\psi-\int_{\partial V}\sqrt{\left|h\right|} ~n_\mu {\partial\mathcal{L}_\text{matter}\over\partial\nabla_\mu\psi} ~ \delta\psi \quad,
\label{eq:delta_full_action}
\end{align}
where $\mathcal{L}_\text{matter}(\psi,\nabla_\mu\psi)$ denotes the Lagrangian density of the matter fields residing in $V$. The bulk terms in the above expression vanish because they are proportional to the equations of motion, which are satisfied by the metric and matter fields order by order in perturbation theory. Hence, $\delta^{\left(n\right)}S_{\textrm{full}}$ corrections receive contributions only from the boundary terms. This is not surprising, because we evaluated the change in the value of the classical action on-shell as the boundary conditions are varied.  The linear order correction is given by\footnote{The boundary term associated with the matter fields starts at quadratic order, since the matter fields in our analysis represent a small perturbation of the Ricci-flat Schwarzschild background.} 
\begin{align}
\label{eq:d1S_full}
\delta^{(1)} S_{\textrm{full}}=\frac{1}{16\pi}\int_{\partial V}\sqrt{\left|h\right|}\left(h^{\mu\nu}\mathcal{K}-\mathcal{K}^{\mu\nu}\right)  g^{(1)}_{\mu\nu} ,
\end{align}
where the extrinsic curvature of \(\partial V\) is evaluated in the unperturbed metric $g_{\mu\nu}^S$. Hence, $\delta^{(1)} S_{\textrm{full}}$ is completely determined by the boundary conditions and the unperturbed metric $g_{\mu\nu}^S$. Although this term does not necessarily vanish, it does not encode any information about the coupling between the induced polarization and external perturbations. Consequently, it is not associated with \eqref{eq:PP_action_weak_fields} and should be subtracted.

Expanding the full action to second order yields
\begin{align}
\label{eq:d2S_full}
\delta^{\left(2\right)}S_{\textrm{full}} &=\frac{1}{8\pi}\int_{\partial V}\sqrt{\left|h\right|}\left(h^{\mu\nu}\mathcal{K}-\mathcal{K}^{\mu\nu}\right)  g^{(2)}_{\mu\nu}
+\frac{1}{16\pi}\int_{\partial V}\sqrt{\left|h\right|}\left(\frac{1}{2}h^{\alpha\beta}h^{\mu\nu}\mathcal{K}-\frac{1}{2}\mathcal{K}^{\mu\nu}h^{\alpha\beta}\right) g^{(1)}_{\alpha\beta} g^{(1)}_{\mu\nu} \notag \\
& +\frac{1}{16\pi}\int_{\partial V}\sqrt{\left|h\right|}\left(h^{\mu\nu}\delta^{(1)} \mathcal{K}-\mathcal{K}\, h^{(1)\mu\nu}-g^{\mu\alpha}g^{\beta\nu}\delta^{(1)} \mathcal{K}_{\alpha\beta}+2\mathcal{K}^{\mu\alpha}g^{\nu\beta} g^{(1)}_{\alpha\beta}\right) g^{(1)}_{\mu\nu} \notag\\
&-\int_{\partial V}\sqrt{\left|h\right|} ~n_\mu {\partial\mathcal{L}_\text{matter}\over\partial\nabla_\mu\psi}\Big|_{\psi=\psi^{(1)}} ~ \psi^{(1)}
\end{align}
where the variation of the extrinsic curvature is given in Appendix~\ref{app:curvature_variations}, and the unperturbed geometric quantities are evaluated in the Schwarzschild metric \eqref{eq:Schw}. Most of the terms in the above expression are completely determined by $g_{\mu\nu}^S$ and by the boundary value $\delta g_{\mu\nu}|_{r=R}$, which is identified with a prescribed weakly curved background. These contributions do not capture the coupling of the polarized Schwarzschild black hole to external perturbations and are therefore not part of the effective action \eqref{eq:PP_action_weak_fields}. The terms of interest are those involving variations of the extrinsic curvature. We analyze them in Section \ref{sec:gravito-magnetic}.

Finally, the expansion \eqref{eq:action_perturbation_def} admits a thermodynamic interpretation. In the classical limit, the free energy \(F\) is related to the full gravitational action via \cite{Gibbons:1977}
\begin{align}
\label{eq:free_energy_is_action}
\beta F=-iS_{\text{full}}\Big|_{\underset{\int dt=-i\beta}{\text{on-shell}}} +\ldots
\end{align}
where the inverse temperature $\beta=1/T$ is identified with the periodicity of Euclidean time, and the ellipsis denotes quantum corrections, which are neglected in the classical regime. We emphasize that there are no conserved charges or associated chemical potentials in our setup. We impose the standard Dirichlet boundary conditions on the metric and matter fields, so the path integral computes the canonical partition function. The energy, given by the expectation value of the Hamiltonian, explicitly tracks the total energy of the spacetime, including the black hole energy, the background energy, and the interaction energy between the background and the black hole.

In the case of the unperturbed Schwarzschild metric \eqref{eq:Schw}, both $S_\text{matter}$ and the Ricci scalar $\mathcal{R}$ vanish, and therefore the free energy is entirely determined by the GHY term in \eqref{eq:full_exact_action}. In this case, the outward-pointing normal vector to $\partial V$ is $n^\mu=\sqrt{f}\,\delta^\mu_r$. The non-vanishing components of the extrinsic curvature are then
\begin{equation}
\label{eq:K_Sch}
    \mathcal{K}^{tt}_S=- \partial_r f^{-1/2}\Big|_{r=R} ~, \quad \mathcal{K}^{ij}_S=-{\sqrt{f}\over R^3} ~\Omega^{ij} \quad \Rightarrow \quad \mathcal{K}_S={d-2\over R} \sqrt{f} + {1\over 2} {\partial_rf\over \sqrt{f}}\Bigg|_{r=R} \quad .
\end{equation}
Subtracting $\mathcal{K}_0=(d-2)/R$, which corresponds to the extrinsic curvature of $\partial V$ embedded in flat space, and evaluating the GHY term in the limit $R\to \infty$ yields
\begin{equation}
    F_\text{Sch}={\Omega_{d-1}\over 16\pi} r_s^{d-3}={m\over d-2} ~.
\end{equation}

A weak external perturbation of the Schwarzschild black hole by a matter field, $\delta\psi$, yields the following correction to the black hole free energy, including terms that will be subtracted later
\begin{equation}
\label{eq:free_energy_is_action2}
    \beta F = \beta F_\text{Sch}+{i\over 2}\int_{\partial V}\sqrt{\left|h\right|} ~n_\mu {\partial\mathcal{L}_\text{matter}\over\partial\nabla_\mu\psi}\Big|_{\psi=\psi^{(1)}} ~ \psi^{(1)}-\frac{i}{16\pi}\int_{\partial V}\sqrt{\left|h\right|}\left(h^{\mu\nu}\mathcal{K}-\mathcal{K}^{\mu\nu}\right)  g^{(2)}_{\mu\nu}
    + \mathcal{O}(\delta\psi^4)~.
\end{equation}
To obtain this expression, we set $g_{\mu\nu}^{\left(1\right)}=0$ in \eqref{eq:d2S_full}. The last term, along with certain components of the remaining correction discussed in the following sections, should be subtracted, as they are entirely determined by the unperturbed metric $g_{\mu\nu}^S$ and the prescribed boundary values of the fields, which correspond to a fixed weakly curved background. These contributions are not related to the free energy associated with the induced response of a Schwarzschild black hole to external perturbations and are therefore subtracted and not included in the free-energy correction.

As we show in the following sections, combining these ideas with those of subsection \ref{sec:GRstresstensor} leads to the polarization of a Schwarzschild black hole induced by external perturbations.

\section{Generalized Smarr relation}
\label{sec:smarr}

The first law of thermodynamics for an asymptotically flat Schwarzschild black hole takes the simple form \cite{Bardeen:1973gs}
\begin{align}
\delta U=T \delta S,
\label{eq:FirstLaw}
\end{align}
where $T=\beta^{-1}$ denotes the temperature and $S$ the black hole entropy. This relation is valid only for an isolated black hole. In the presence of external fields, however, the black hole develops an induced polarization, the total energy of the system is modified, and the above form of the first law must be generalized.

A useful analogy is provided by linear materials subjected to external electromagnetic fields. In the absence of an applied field, microscopic moments are either absent or randomly oriented, yielding no net polarization. An external field couples to these internal degrees of freedom and partially aligns them, inducing a macroscopic polarization. This process reorganizes the microscopic state of the system, reduces orientational disorder, and competes with thermal fluctuations. Consequently, the application of an external field modifies the entropy and alters the thermodynamic potentials.

A similar physical mechanism governs the response of black holes. Any perturbing background field plays a role similar to that of an electromagnetic field in linear media, since gravity universally couples to all forms of energy. To account for this phenomenon within black hole thermodynamics, we introduce a pair of conjugate variables: $\mathcal{M}$, a tensor characterizing the multipole structure of the external perturbation, and $\mathcal{P}$, the polarization tensor induced on the Schwarzschild black hole. For notational simplicity, tensor indices are suppressed. The first law \eqref{eq:FirstLaw} is then supplemented by an additional term of the form
\begin{align}
\label{eq:Schw-1stLaw}
\delta U = T \delta S - \mathcal{P} \delta \mathcal{M}.
\end{align}

The form of this correction is completely determined by the definition of the free energy in \eqref{eq:free_energy_is_action2}. In our setup, the path integral computes $F(T,\mathcal{M})$, since the Dirichlet boundary conditions for the perturbations are specified entirely by the external background field $\mathcal{M}$. Consequently,
\begin{equation}
\delta F = -S\,\delta T - \mathcal{P}\,\delta \mathcal{M} ~.
\end{equation}
Performing a Legendre transform that replaces the temperature $T$ with its conjugate $S$ yields the above $U(S,\mathcal{M})$, where the internal energy is identified with the expectation value of the Hamiltonian generating time translations of the entire system.

Strictly speaking, our treatment establishes the extension of the standard thermodynamic relations only at leading order in the perturbative expansion. For many of the calculations presented below, however, extending the analysis to higher orders is straightforward. A fully non-perturbative treatment, by contrast, may be obstructed by ambiguities, and we do not propose a framework for a non-perturbative definition of black hole polarization.

Since Love numbers characterize the linear response to external perturbations, they naturally play a central role in determining the induced polarization of the black hole and, as will be shown below, enter the extended thermodynamic description.

Although the actions of the systems considered in this paper are not scale invariant, their equations of motion admit a scaling symmetry: they remain invariant under rescalings of the coordinates, provided the metric and other fields are held fixed. Such transformations map solutions of the equations of motion into new, physically inequivalent solutions. Equivalently, a family of solutions can be generated by appropriately rescaling the dimensionful parameters of a given configuration. The Smarr relation is obtained by exploiting this scaling symmetry to integrate the generalized first law \eqref{eq:Schw-1stLaw}.

By dimensional analysis, a coordinate rescaling $x^\mu\to \lambda x^\mu$ induces the following transformations of the thermodynamic variables
\begin{align}
\label{eq:gravity_is_scale_free}
U\rightarrow\lambda^{d-3}U,\,\,\,S\rightarrow\lambda^{d-2}S,\,\,\,\mathcal{M}\rightarrow\lambda^{-l}\mathcal{M},\,\,\,\mathcal{P}\rightarrow\lambda^{d-3+l}\mathcal{P} ~.
\end{align}
Setting $\lambda = 1+\varepsilon$, with $\varepsilon \ll 1$, yields
\begin{align}
\label{eq:thermodynamic_variation}
\delta U=\left(d-3\right)\varepsilon U,\,\,\,\delta S=\left(d-2\right)\varepsilon S,\,\,\, \delta\mathcal{M}=-l\varepsilon\mathcal{M},
\,\,\,\delta \mathcal{P} =\left(d-3+l\right)\varepsilon \mathcal{P} ~.
\end{align}
Substituting these variations into the extended first law \eqref{eq:Schw-1stLaw} gives the generalized Smarr relation \cite{Smarr:1972kt},
\begin{align}
\label{eq:TS_U_QP}
(d-3)U = (d-2)TS+l\mathcal{M} \mathcal{P}.
\end{align}
Finally, using the free energy \(F=U-TS\), this relation can be rewritten as
\begin{align}
\label{eq:QP_F_U}
\mathcal{M} \mathcal{P}=\frac{(d-2)F-U}{l}.
\end{align}
This relation will be used to extract the polarization tensor of a Schwarzschild black hole in the presence of scalar, electric, gravito-magnetic, and gravito-electric external fields. In the above equation, the multipole $\mathcal{M}$ is entirely specified by the external perturbation, while the prescriptions for computing the internal energy $U$ and the free energy $F$ are given in subsections~\ref{sec:GRstresstensor} and~\ref{sec:exact_gravitational_action}, respectively.

\section{Warm-up example: conducting sphere}
\label{sec:example}

In this section, we illustrate the general idea behind deriving an effective action directly from a microscopic theory by working through a simple and familiar example from classical electrodynamics. We begin by reviewing the standard approach, which relies on Feynman diagrams and proceeds by matching the exact field amplitude to its counterpart in the effective field theory in order to determine the Wilson coefficients of the effective action. We then introduce an alternative method in which the effective action is obtained directly by evaluating the full action on shell, without relying on Feynman diagrams. This approach is manifestly gauge invariant, as it amounts to evaluating the action – a gauge-invariant quantity – and does not require fixing a gauge on the EFT side.

\begin{figure}[t!]
\centering \noindent
\includegraphics[width=6cm]{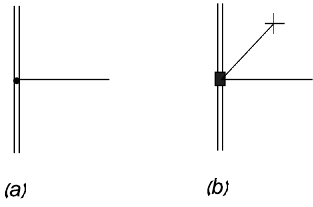}
\caption[]{Feynman diagrams contributing to the asymptotic value of
$\mathcal{A}_0$ in the EFT approach (\ref{EMphi_eff_action}). The dot and the heavy box denote insertions of the $q\mathcal{A}_0$ and
$-2\alpha\,(\nabla\overline{\mathcal{A}}_0 \cdot \nabla\mathcal{A}_0)$ vertices, respectively. In figure (b), the line with a cross at the end represents the background field $\overline{\mathcal{A}}_0$, while the solid line denotes the propagator \eqref{eq:A0propagator}.}
\label{EMphi_EFT_diag}
\end{figure}

Consider the effective point-particle action describing the interaction of an electromagnetic field with a perfectly conducting sphere of radius $r_s$. Assuming that the characteristic spacetime scales associated with the electromagnetic field are large compared to $r_s$, the effective action that respects gauge and reparameterization invariance takes the form \cite{Goldberger:2004jt} 
 \begin{equation}
 S_\mt{EFT}[\mathcal{A}_{\mu}] = -\frac{1}{4}\int d^4x F_{\mu\nu}F^{\mu\nu} -q\int dx^{\mu}\mathcal{A}_{\mu}- \alpha\int d\tau E_{\mu}E^{\mu}
 - \beta\int d\tau B_{\mu}B^{\mu} +\ldots,
 \end{equation}
where $\alpha$ and $\beta$ are the electromagnetic analogs of the Love numbers introduced in subsection~\ref{sec:BH-EFT}, and $E_{\mu}, B_{\mu}$ denote the electric and magnetic components of the Maxwell field strength tensor,
 \begin{align}
 \label{eq:E_mu def}
 E_{\mu}&=F_{\mu\nu} {d x^{\nu} \over d \tau}~,
 \non
 B_{\mu}&= {d x^{\alpha} \over d \tau} \epsilon_{\alpha\mu\beta\gamma}F^{\beta\gamma}~.
 \end{align}
These components are the electromagnetic analogs of \eqref{eq:gravito_EM_components_E} and \eqref{eq:gravito_EM_components_B}. Decomposing the electromagnetic field $\mathcal{A}_{\mu}$ into spatial and
temporal parts $\mathcal{A}_{\mu}\rightarrow (\mathcal{A}_0,\mathcal{A}_I)$ and assuming the problem is static yields
\footnote{This decomposition is analogous to \eqref{KK} in GR. In the static case, time-reversal symmetry $t\rightarrow -t$ implies $\mathcal{A}_I=0$, and consequently $B_{\mu}=0$ as well. }
 \begin{align}
 S_\mt{EFT}[\mathcal{A}_0] &= \frac{1}{2}\int d^4 x (\nabla\mathcal{A}_0)^2 -q\int dt \, \mathcal{A}_0 + \alpha \int dt \, \big(\nabla\mathcal{A}_0\big)^2
 +\cdots.
 \label{eq:EM_eff_action}
 \end{align}
As the problem is time-independent, the time integral, $\int dt$, factors out and will be suppressed for the remainder of this section. The propagator of the electrostatic potential is given by Coulomb’s law
 \begin{equation}
 \langle \mathcal{A}_0(\textbf{x})\mathcal{A}_0(\textbf{y}) \rangle= 
 {1 \over 4\pi|\textbf{x} - \textbf{y}|}~.
 \label{eq:A0propagator}
 \end{equation}
We are interested in matching the coefficients $q$
and $\alpha$. To this end, let us introduce the following
electrostatic potential
 \begin{equation}
 \overline{\mathcal{A}}_0=-\bf{E} \cdot \bf{x}\, \label{scalar_background},
 \end{equation}
where $\bf{E}$ is a constant electric field.
When the above background is introduced into the effective action (\ref{eq:EM_eff_action}), it takes the following form
 \begin{align}
 S_\mt{EFT}[\overline{\mathcal{A}}_0+\mathcal{A}_0] &= \frac{1}{2}\int d^{\,3}x (\nabla\overline{\mathcal{A}}_0)^2+\int d^{\,3}x \nabla\overline{\mathcal{A}}_0 \cdot \nabla\mathcal{A}_0
 +\frac{1}{2}\int d^{\,3}x \big(\nabla\mathcal{A}_0\big)^2 \non
 &-q\mathcal{A}_0 + \alpha\,(\nabla\overline{\mathcal{A}}_0)^2+2\,\alpha\,(\nabla\overline{\mathcal{A}}_0 \cdot \nabla\mathcal{A}_0)
 +\alpha\,(\nabla\mathcal{A}_0)^2
 +\cdots,
 \label{EMphi_eff_action}
 \end{align}
where all terms in the second line are evaluated at the location of the point-like conducting sphere. The diagrams that contribute to the asymptotic value of the electrostatic potential are shown in Fig. \ref{EMphi_EFT_diag}
 \begin{align}
 \text{Fig.}\ref{EMphi_EFT_diag}\text{(a)}&={q \over 4\pi|\bf{x}|}\non
 \text{Fig.}\ref{EMphi_EFT_diag} \text{(b)}&= - {\alpha\over 2\pi}\,\int d^{\,3}\textbf{y} \, \frac{\textbf{E} \cdot \nabla\delta(\textbf{y})}{|\bf{x} - \bf{y}|}
 = {\alpha\over 2\pi}\,\frac{\bf{E} \cdot \bf{x}}{|\bf{x}|^{3}}
 \end{align}
Thus, the amplitude of the electrostatic potential in the EFT approach is given by
 \begin{equation}
 \mathcal{A}_0^\mt{EFT}(\textbf{x})={q \over 4\pi|\bf{x}|}-\textbf{E} \cdot \textbf{x}\Big( 1- {\alpha \over 2\pi|\textbf{x}|^{\,3}}\Big) \, .
 \end{equation}
On the other hand, the exact solution to the Laplace equation that satisfies the boundary conditions at infinity and at $r=r_s$ can be written as follows
 \begin{equation}
 \mathcal{A}_0^\mt{full}(\textbf{x})={Q \over 4\pi|\textbf{x}|}-\textbf{E} \cdot \textbf{x} ~ F_\mt{EM}(r) ~, \quad F_\mt{EM}(r) = \Big( 1- \Big({r_s \over r}\Big)^3\Big)
 \, \label{phi_full},
 \end{equation}
where $Q$ is the total charge of the sphere. Matching $\mathcal{A}_0^\mt{full}$ and $\mathcal{A}_0^\mt{EFT}$ then yields \cite{Goldberger:2004jt}
 \begin{equation}
 q=Q~,  \quad \alpha=2\pi \, r_s^{\,3}\, .
 \label{constants}
 \end{equation}

The method presented above uses the underlying symmetry structure, together with the matching of physical observables, to construct the point-particle terms of the effective action \eqref{eq:EM_eff_action} and determine the values of the unknown parameter $q$ and $\alpha$. We now illustrate the idea of deriving the point-particle effective action directly from the microscopic action. This approach offers several advantages, most notably the elimination of the need to compute Feynman diagrams.

The action of the full theory is given by,
\begin{equation}
    S_\mt{full}[\mathcal{A}_{\mu}] = \frac{1}{2}\int_V  \mathcal{A}_\nu \partial_\mu F^{\mu\nu}+{1\over 2}\int_{\partial V} \sqrt{|h|}\,n_\mu\,  F^{\mu\nu}\mathcal{A}_\nu - \int_V  J^\mu \mathcal{A}_\mu ~.
\end{equation}
Here, $J^\mu$ is a conserved current, and $\partial V$ is defined as in subsection~\ref{sec:exact_gravitational_action}. The surface term is analogous to the GHY boundary term in gravity, and is chosen such that, for fields $A_\mu$ satisfying the equations of motion, the action $S_\mt{full}$ is stationary under variations of $A_\mu$ that vanish on $\partial V$. Without this term, the presence of second derivatives in the bulk would render the variational principle ill-defined. This representation of $S_\mt{full}$ is analogous to the full gravitational action \eqref{eq:full_exact_action}. In the static limit, this reduces to
 \begin{equation}
 S_\mt{full}[\mathcal{A}_0] = \frac{1}{2}\int d^3\textbf{x}\, \mathcal{A}_0 (-\Delta\mathcal{A}_0) 
 + \frac{R^2}{2}\int \mathcal{A}_0 \,\partial_r\mathcal{A}_0 \, d\Omega
 -\int d^3\textbf{x} \, \rho \, \mathcal{A}_0
 ~,
 \label{EMfull_action}
 \end{equation}
where the integral over time has been omitted, $\rho=J^0$ is the charge density, and $d\Omega$ is the volume element on the unit sphere.

In the direct approach, the background field (\ref{scalar_background}) represents the long-wavelength mode and serves as the boundary condition,
\begin{equation}
 \mathcal{A}_0^\mt{full}(\textbf{x})|_{|\textbf{x}|=R\gg r_s} = \overline{\mathcal{A}}_0 ~,
\end{equation}
where the {\it finite} scale $R\gg r_s$ represents the domain where the EFT applies. Additionally, a boundary condition must be imposed on the surface of the conducting sphere. For simplicity, we assume the sphere is grounded, so that
 \begin{equation}
 \mathcal{A}_0^\mt{full}(\textbf{x})\Big|_{|\textbf{x}|=r_s}=0 \quad .
 \end{equation} 
Integrating out the short-wavelength modes to obtain the effective action $S_\mt{EFT}$ amounts to evaluating the full action $S_{\textrm{full}}$ on the classical solution $\mathcal{A}_0^\mt{full}$ of the equations of motion, with the long-wavelength components acting as boundary data,
\begin{equation}
    S_\mt{EFT}[\overline{\mathcal{A}}_0] = S_\mt{full}[\mathcal{A}_0^\mt{full}]~.
\end{equation}

The electrostatic potential $\mathcal{A}_0^\mt{full}$ that satisfies the Laplace equation and the above boundary conditions is given by
 \begin{equation}
 \mathcal{A}_0^\mt{full}(\textbf{x})=-{\textbf{E} \cdot \textbf{x}\over F_\mt{EM}(R)} ~ F_\mt{EM}(r)
 ~.
 \end{equation}
This solution coincides with (\ref{phi_full}) in the limit $R\rightarrow\infty$. Hence, the full action (\ref{EMfull_action}) evaluated on-shell takes the form
\begin{align}
 S_\mt{EFT}[\overline{\mathcal{A}}_0] = S_\mt{full}[\mathcal{A}_0^\mt{full}] &= \frac{R^2}{2 F_\mt{EM}(R)}\int (\textbf{E} \cdot \textbf{x}) \partial_r \Big(\textbf{E} \cdot \textbf{x} ~ F_\mt{EM}(r)\Big) \cdot d\Omega 
 \nonumber \\
 &= \frac{R^2}{2 }\int (\textbf{E} \cdot \textbf{x}) \partial_r(\textbf{E} \cdot \textbf{x}) \cdot d\Omega 
 + \frac{F'_\mt{EM}(R)}{2 F_\mt{EM}(R)} \, R^2\int (\textbf{E} \cdot \textbf{x})^2 d\Omega ~.
\end{align}
The first term in the second line represents the background contribution to the effective action and does not encode any information about the conducting sphere, whereas the second term captures the interaction of the sphere with the long-wavelength background field. Evaluating these contributions gives
 \begin{equation}
 S_\mt{EFT}[\overline{\mathcal{A}}_0] = {4\pi\over 6}\textbf{E}^2 R^3 +2\pi\textbf{E}^2 r_s^3 \Big( 1 + \mathcal{O}(r^3_s/R^3) \Big)={1\over 2}\textbf{E}^2 V + 2\pi r_s^{\,3}(\nabla\overline{\mathcal{A}}_0)^2 \, .
 \end{equation}
Thus, we recovered the effective action \eqref{eq:EM_eff_action} with 
 \begin{equation}
 \alpha=2\pi \, r_s^{\,3}\, ,
 \end{equation}
in full agreement with \eqref{constants}.

\section{Scalar background}
\label{sec:scalar_bckgrnd}

For a minimally coupled scalar field $\psi$, the matter action in \eqref{eq:full_exact_action} is
\begin{equation}
S_\text{matter}= {1\over 2} \int_V \sqrt{|g|}\, g^{\mu\nu}\partial_\mu\psi\partial_\nu\psi = - {1\over 2} \int_V \sqrt{|g|}\,\psi\Box\psi - {1\over 2} \int_{\partial V}\sqrt{\left|h\right|} ~n_\mu \psi \partial^\mu\psi~.
\label{eq:min_scalar_action}
\end{equation}
The $d$-dimensional Schwarzschild background \eqref{Sch metric} with $\psi=0$ is an exact solution of the equations of motion. At distances much larger than $r_s$, the effective action \eqref{eq:PP_action} receives additional contributions that encode the coupling of the black hole to long-wavelength scalar field modes\footnote{We retain only curvature-independent quadratic terms in $\psi$, and accordingly do not distinguish between different index contractions in \eqref{eq:scalarSpp}.}
\begin{multline}
    \label{eq:scalarSpp}
    S_\text{p.p.} = \sum_{l=1}^\infty\frac{(-1)^l\lambda_{l}}{2l!}\int d\tau \nabla_{\mu_1}\ldots \nabla_{\mu_l}\psi\,\nabla^{\mu_1}\ldots\nabla^{\mu_l}\psi  +\ldots
    \\
    =\sum_{l=1}^\infty\frac{\lambda_{l} }{2l!} \, e^{(2\,\hat l+1)\phi}\gamma^{I_1J_1}\cdots\gamma^{I_lJ_l}\,\nabla_{I_1}\ldots \nabla_{I_l}\psi(0)\,\nabla_{J_1}\ldots\nabla_{J_l}\psi(0)\int dt  +\ldots ~,
\end{multline}
where $\nabla_\mu$ and $\nabla_I$ denote the covariant derivatives compatible with $g_{\mu\nu}$ and $\gamma_{IJ}$, respectively. The remaining contributions, including those appearing in \eqref{eq:PP_action}, are absorbed into the ellipses and will not be considered here. We now derive this action from \eqref{eq:action_perturbation_def}.

Consider a scalar configuration carrying a multipole number $l$,\footnote{For the remainder of this section, Latin indices are raised and lowered using the Kronecker delta $\delta_{IJ}$.}
\begin{equation}
\label{eq:scalar_bckrnd}
\overline{\psi}=\Psi_{J_{1} \cdots J_{l}}x^{J_{1}}...x^{J_{l}} ~, \quad
|\Psi_{J_{1} \cdots J_{l}}|r_s^l\ll 1 ~ .
\end{equation} 
where $\Psi_{J_{1} \cdots J_{l}}$ is a totally symmetric and traceless tensor. In flat spacetime this profile is an exact solution of the equations of motion for $\psi$. In the presence of the Schwarzschild black hole, however, it is modified as follows \cite{Kol:2011vg,Hui:2020xxx}
\begin{equation}
    \delta\psi = \overline{\psi} {F_\mt{S}(r)\over F_\mt{S}(R)}    ~, \quad  \delta\psi\Big|_{r=R}=\overline{\psi} ~,
    \label{eq:scalar_full}
\end{equation}
where regularity is imposed at the horizon $r=r_s$, and
\begin{equation}
 F_\mt{S}(r)={\Gamma^2(1+\hat{l}) \over \Gamma(1+2\hat{l})}
 \;{}_2F_1\Big( - \hat{l} ~,~ -\hat{l} ~,~1~;1-\Big({r_s\over r}\Big)^{d-3}\Big)
 ~, \quad \hat{l}={l\over d-3} ~.
 \label{eq:FS}
\end{equation}
This solution is valid at linear order in the weak perturbation. Beyond leading order, the backreaction on the geometry must be taken into account, since the configuration \eqref{eq:scalar_full} generates a weak curvature near the black hole via its stress–energy tensor.

Substituting \eqref{eq:scalar_full} into \eqref{eq:d2S_full} and using the fact that there are no linear corrections to the Schwarzschild metric, $g^{(1)}_{\mu\nu}=0$,  we obtain
\begin{equation}
    \delta S_\text{full} = 
    \frac{1}{16\pi}\int_{\partial V}\sqrt{|h_S|}\left(h^{\mu\nu}_S\mathcal{K}_S-\mathcal{K}_S^{\mu\nu}\right) g^{(2)}_{\mu\nu}
     - { f(R)\over 2 } \int_{\partial V}  \, \overline{\psi} \partial_r \overline{\psi}
     - { f(R)\over 2 }{F'_\mt{S}(R)\over F_\mt{S}(R)} \int_{\partial V}  \, \overline{\psi}^2 
     + \mathcal{O}(\overline{\psi}^4)~.
\end{equation}
The effective action \eqref{eq:action_perturbation_def}, describing the black hole coupled to the long-wavelength scalar modes $\overline{\psi}$, is obtained by subtracting the terms in the above expression that depend solely on the unperturbed metric and the fixed weakly curved background – for instance, the first two terms appearing in the above expression. Hence, the effective action to second order in the scalar perturbation is
\begin{equation}
    S_{\textrm{EFT}}=S_\mt{Sch} - {f(R)\over 2 }{F'_\mt{S}(R)\over F_\mt{S}(R)} \int_{\partial V}  \, \overline{\psi}^2 ~,
\end{equation}
Using \eqref{use_iden1} and \eqref{use_iden2} to evaluate the spherical integral, gives
\begin{equation}
    S_{\textrm{EFT}}=S_\mt{Sch} - { R^{2l+d-2}\over 2 }{F'_\mt{S}(R)\over F_\mt{S}(R)}f(R) 
     {\Omega_{d-1+2l}\over (2\pi)^l} ~ l! \int dt \, 
 \Psi_{J_{1} \cdots J_{l}} \Psi^{J_{1} \cdots J_{l}}~.
\end{equation}
Expanding this expression in the limit $R\gg r_s$ and retaining only the constant term yields
\begin{equation}
    S_{\textrm{EFT}}=S_\mt{Sch}+\,r_s^{2l+d-3}{ (d-3)\pi \over 2 (16)^{\hat l} } 
     {\Gamma(2\hat l) \Gamma(1-2\hat l)\over \Gamma^2(-\hat l)\Gamma^2\big({1\over 2}+\hat l\big)}
     {\Omega_{d-1+2l}\over (2\pi)^l} ~ l! \int dt\, 
 \Psi_{J_{1} \cdots J_{l}} \Psi^{J_{1} \cdots J_{l}} + \ldots ~.
\end{equation}
Only the contribution arising from the polarization of the black hole induced by the external scalar field is displayed.\footnote{Scheme-dependent terms, namely those that depend explicitly on $R$, are subtracted. These terms are sensitive to the shape of the boundary surface and to its rescalings, and most of them vanish in the limit $R \to \infty$. Only a small subset remains unsuppressed by inverse powers of $R$. These residual contributions correspond to the self-energy of the background and to the interaction energy between the unperturbed Schwarzschild black hole and the background. In contrast, quantities associated solely with the thermodynamics of the black hole are independent of the boundary scale $R$. Consequently, the $R$-dependent terms do not encode information about the induced polarization of the black hole and are therefore omitted from the analysis. \label{foot:scheme-dependent}}

The above result can be written in the compact form
\begin{equation}
    \delta S_{\textrm{EFT}}= {\lambda_l \over 2 l!} \int dt ~ 
    \partial_{J_1}...\partial_{J_l}\overline{\psi} \, \partial^{J_1}...\partial^{J_l}\overline{\psi} ~,
    \label{eq:Scalar_eff_action}
\end{equation}
where $\lambda_l$ is the Love number, in agreement with previous calculations \cite{Kol:2011vg,Hui:2020xxx}.
\begin{equation}
     \lambda_l=
     {\Omega_{d-1+2l}\over (2\pi)^l}
     { (d-3) \over 2 (16)^{\hat l} } 
     {\Gamma^2(1+\hat l) \over \Gamma^2({1\over 2}+\hat l)}\tan(\pi \,\hat l)
      \, r_s^{2l+d-3} ~.
\end{equation}

\subsection{Polarization Thermodynamics}
In this subsection, we study the thermodynamic aspects of Love numbers based on the results derived in subsections \ref{sec:GRstresstensor}, \ref{sec:exact_gravitational_action} and \ref{sec:smarr}. 

In the presence of the static scalar background \eqref{eq:scalar_bckrnd}, the black hole effective action, given in \eqref{eq:PP_action} and 
\eqref{eq:scalarSpp}, takes the form
\begin{multline}
S_\text{p.p.}=-m \int dt \, \phi+
\sum_{l=1}^\infty\frac{\lambda_{l}}{l!} \int dt \, \partial_{I_1}\ldots \partial_{I_l}\psi ~ \partial^{I_1}\ldots\partial^{I_l}
\overline\psi
\\
+\sum_{l=1}^\infty\frac{\lambda_{l}}{2 l!} \int dt
\Big(-l\sigma^{I_1J_1} 
+ (1+2\hat l)\phi\,\delta^{I_1 J_1}\Big)
\partial_{I_1}\partial_{I_2}\ldots\partial_{I_l}\overline\psi ~ \partial_{J_1}\partial^{I_2}\ldots\partial^{I_l}\overline\psi + \ldots ~,
\label{eq:Spp_scalar}
\end{multline}
where we decompose the scalar field into a background and an induced perturbation, $\psi \to \overline{\psi} + \psi$, and retain only terms linear in $\phi$, $\sigma_{IJ}$, and $\psi$. The matter action is given by
\begin{equation}
 S_\text{matter}=- {1\over 2}\int_V\,\sqrt{\gamma}\,\gamma^{IJ} \partial_I\psi\partial_J\psi \quad . 
\end{equation}
Hence, the propagator of the scalar field is
\begin{equation}
\label{eq:propPsi}
    \left\langle \psi\left(x\right)\psi\left(0\right)\right\rangle =\frac{-1}{2\pi\,\Omega_{d-3}} \, {1\over |x|^{d-3}} \quad .
\end{equation}

As shown in subsection \ref{sec:GRstresstensor}, the first term in \eqref{eq:Spp_scalar} accounts for the energy of the unperturbed Schwarzschild black hole, given in \eqref{eq:USch}. The background configuration \eqref{eq:scalar_bckrnd} does not produce linear corrections to this energy, $U^{(1)}=0$, since the energy–momentum tensor that sources the gravitational fields is quadratic in the scalar perturbation. 

\begin{figure}[t!]
\centering \noindent
\includegraphics[width=11cm]{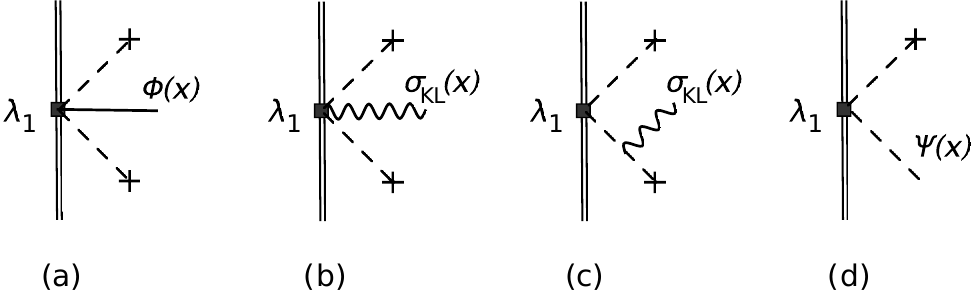}
\caption[]{Feynman diagrams contributing to the asymptotic values of the gravitational fields $\phi$, $\sigma_{KL}$, and $\psi$ in the EFT approach. The black hole worldline is represented by a double line. Dashed lines denote $\psi$ propagators, while a dashed line with a cross represents the background field $\overline{\psi}$. The heavy boxes in panels (a) and (b) correspond to vertices associated with the two terms in the second line of \eqref{eq:Spp_scalar}, whereas those in panels (c) and (d) correspond to the second term in \eqref{eq:Spp_scalar}.}
\label{Fig:BHenergyS}
\end{figure}

The leading correction to $U_\text{Sch}$ therefore arises at quadratic order in $\overline{\psi}$. This contribution is computed below, restricting the analysis to the case $l=1$ for simplicity. All relevant diagrams are shown in Fig.~\ref{Fig:BHenergyS}(a)-(c). Using \eqref{eq:propPhi}, \eqref{eq:propSigma}, \eqref{eq:Spp_scalar}, and \eqref{eq:propPsi}, one obtains
\begin{align}
  &   \text{Fig.}\ref{Fig:BHenergyS}\text{(a)}= 
 \lambda_{1} \frac{2(d-1)}{\left(d-2\right)\Omega_{d-3}}
 ~ \frac{\Psi_{J} \Psi^{J}}{|x|^{d-3}}~,
 \nonumber\\
  &  \text{Fig.}\ref{Fig:BHenergyS}\text{(b)}=
\frac{-8 \lambda_1}{\Omega_{d-3}|x|^{d-3}}\left(\Psi_{K} \Psi_L
-\frac{\delta_{KL}}{d-3} \Psi_{J} \Psi^{J} \right) ~,
\nonumber\\
    &\text{Fig.}\ref{Fig:BHenergyS}\text{(c)}= 
    \frac{8 \lambda_{1}}{\Omega_{d-3}|x|^{d-1}}
    \Big(\Psi_K\Psi_L x^2- (d-3) \Psi_Jx^J\Psi_{(K} x_{L)}\Big) ~,
\\
&\text{Fig.}\ref{Fig:BHenergyS}\text{(d)}= \frac{\lambda_{1}}{\Omega_{d-1}} {\overline\psi\over|x|^{d-1}}~.
    \nonumber
\end{align}
Calculating these diagrams in position space is straightforward. For most of them, no computation beyond the Feynman rules is required. Fig. \ref{Fig:BHenergyS}(c) is the only diagram that involves a single integral over a bulk point. This integral is evaluated in Appendix \ref{apx:integrals}. As a consistency check, one may verify that the above expressions for $\sigma_{KL}$ satisfy the harmonic gauge condition \eqref{eq:GF_action}, namely $\frac{1}{2}\partial_{I}\sigma_{\,J}^{J}=\partial_{J}\sigma_{I}^{\,J}$.

Substituting these expressions into \eqref{eq:U_with_NRG} yields an $R$-independent second-order correction to the unperturbed energy $U_\text{Sch}$. Adding the background energy computed in Appendix \ref{appx:bckgrnd_energy} then gives
\begin{equation}
    U^{(2)}_\infty={\Omega_{d-1}\over 2(d-1)}R^{d-1}\Psi_J\Psi^J-{1\over2} {d-3\over d-1}\lambda_1\Psi_J\Psi^J+\ldots ~,
\end{equation}
where the ellipsis denotes contributions that depend on both $R$ and the black hole mass $m$, and the subscript $\infty$ indicates that this energy corresponds to a field profile $\delta\psi$ that approaches $\overline{\psi}$ at spatial infinity, rather than at finite radius $r = R$ as in \eqref{eq:scalar_full}.

To adapt this result to the boundary condition imposed in our setup – where all fields match the background at finite radius $r = R$ instead of at infinity – it is sufficient to rescale the perturbation amplitude by a constant factor,
\begin{equation}
\Psi_J \to {\Psi_J\over F_\text{S}(R)}~, \quad 
F_\text{S}(R) = 1 +  \frac{\lambda_{1}}{\Omega_{d-1}R^{d-1}} 
+\ldots \quad .
\end{equation}
The expression for $F_\text{S}(R)$ given above follows from the diagram in Fig.~\ref{Fig:BHenergyS}\text{(d)}, and it is consistent with the asymptotic expansion of \eqref{eq:FS}. Additional contributions depend on the black hole mass $m$, but they are not relevant for our purposes. As a result, we obtain
\begin{equation}
    U^{(2)}=-{1\over2} \lambda_1\Psi_J\Psi^J+\ldots ~,
\end{equation}
where the $R$-dependent terms – including those that depend on $m$ – have been suppressed (see footnote \ref{foot:scheme-dependent}).

On the other hand, using \eqref{eq:action_perturbation_def}, \eqref{eq:free_energy_is_action}, and \eqref{eq:Scalar_eff_action}, the free energy associated with the polarization of the black hole is given by
\begin{equation}
F^{(2)} = - \frac{\lambda_1}{2}\Psi_J \Psi^J .
\end{equation}
The second-order correction to $TS$ then reads
\begin{equation}
    \delta^{(2)}(TS)=U^{(2)}-F^{(2)} =0 \quad .
\end{equation}
Substituting these expressions into \eqref{eq:QP_F_U}, we obtain
\begin{equation}
\mathcal{M}_{J}\mathcal{P}^{J}=-\frac{d-3}{2}\lambda_{1}\Psi_{I}\Psi^{I} .
\end{equation}
Identifying $\mathcal{M}_J$ with the external control parameter $-\Psi_J$ then yields the desired polarization vector,
\begin{equation}
    \mathcal{P}^{I} = \frac{d-3}{2}\lambda_{1}\Psi^{I} \quad .
\end{equation}

\section{Maxwell background}
\label{sec:maxwell_bckgrnd}

In this section we consider a minimally coupled Maxwell field, $\mathcal{A}_\mu$, in a general number of spacetime dimensions. The matter action appearing in \eqref{eq:full_exact_action} takes the form
\begin{equation}
\label{eq:Maxwell_action}
    S_\mt{matter}[\mathcal{A}_{\mu}] = \frac{1}{2}\int_V  \mathcal{A}_\nu \partial_\mu \Big(|g|^{1/2}  F^{\mu\nu}\Big)+{1\over 2}\int_{\partial V} \sqrt{|h|}\,n_\mu\,  \mathcal{A}_\nu F^{\mu\nu}  ~.
\end{equation}
The $d$-dimensional Schwarzschild metric \eqref{Sch metric} and the trivial gauge field configuration $\mathcal{A}_\mu=0$ solve the exact Einstein–Maxwell equations of motion. We now introduce a weak, static background of the form\footnote{For the remainder of this section, Latin indices are raised and lowered using the Kronecker delta $\delta_{IJ}$.}
\begin{equation}
\label{eq:Maxwell_bckrnd}
\overline{\mathcal{A}}_0=\mathcal{E}_{J_{1} \cdots J_{l}}x^{J_{1}}...x^{J_{l}} ~, \quad
|\mathcal{E}_{J_{1} \cdots J_{l}}|r_s^l\ll 1  ~.
\end{equation} 
This configuration represents an external electric potential carrying a multipole number $l$. It solves the linearized Einstein–Maxwell equations in flat space and will serve as the boundary condition for the gauge field perturbations of the Schwarzschild spacetime.

Because the Maxwell energy–momentum tensor is quadratic in $\mathcal{A}_\mu$, the background \eqref{eq:Maxwell_bckrnd} does not induce linear corrections to the Schwarzschild metric \eqref{eq:Schw}. Furthermore, $\mathcal{A}_I=0$ due to the $t\to -t$ symmetry, and the remaining component of the vector potential satisfies
\begin{equation}
\partial_I\Big( g^{IJ}_S g^{00}_S \sqrt{|g_S|}F_{J0}\Big)=0
\quad\Rightarrow\quad
\Big(f \, \partial_r^2 + {d-2\over r} \, f \, \partial_r + {1\over r^2} \Delta_{\mathbb{S}^{d-2}}\Big)\mathcal{A}_0 =0 ~,
\end{equation}
where $\Delta_{\mathbb{S}^{d-2}}$ denotes the covariant Laplacian acting on a scalar field on the unit $(d-2)$-sphere. It is convenient to introduce a dimensionless radial coordinate $X$ and separate variables as
\begin{equation}
    X=\Big({r_s\over r}\Big)^{d-3} ~, \quad \mathcal{A}_0(X,\Omega)=X^{-\hat l}\mathcal{A}(X) Y_{lm}(\Omega) ~,
\end{equation}
where $Y_{lm}(\Omega)$ is a scalar spherical harmonic on $\mathbb{S}^{d-2}$. The radial profile $\mathcal{A}(X)$ then satisfies the following ordinary differential equation
\begin{equation}
    (1-X)X\mathcal{A}''(X)-2\hat l(1-X)\mathcal{A}'(X)-\hat l(\hat l+1)\mathcal{A}(X)=0 ~.
\end{equation}
The solution that is regular at the horizon $X=1$ and matches \eqref{eq:Maxwell_bckrnd} at $r=R$ is given by
\begin{equation}
    \mathcal{A}_0(r,\Omega)=\overline{\mathcal{A}}_0 \, {F_\mt{E}(r)\over F_\mt{E}(R)} \quad ,
    \label{eq:vector_full}
\end{equation}
where
\begin{equation}
    F_\mt{E}(r) ={\Gamma(\hat l+2)\Gamma(\hat l)\over 2\Gamma(2\hat l)} \, \Big(1-\Big({r_s\over r}\Big)^{d-3}\Big) 
    \;{}_2F_1\Big( 1- \hat{l} ~,~ -\hat{l} ~,~2~;1-\Big({r_s\over r}\Big)^{d-3}\Big) 
    \quad .
    \label{eq:FE}
\end{equation}
Using \eqref{eq:action_perturbation_def} and \eqref{eq:d2S_full}, together with the fact that there are no linear corrections to the Schwarzschild metric, $g^{(1)}_{\mu\nu}=0$, we obtain
\begin{align}
    S_{\textrm{EFT}}=S_{\textrm{full}}\Big|_\text{on-shell}
    &=S_\mt{Sch} +\frac{1}{16\pi}\int_{\partial V}\sqrt{|h_S|}\left(h^{\mu\nu}_S\mathcal{K}_S-\mathcal{K}_S^{\mu\nu}\right) g^{(2)}_{\mu\nu}
      \\
     &+ { 1\over 2 } \int_{\partial V}  \, \overline{\mathcal{A}}_0 \partial_r \overline{\mathcal{A}}_0
    + {1\over 2}  {F'_\mt{E}(R)\over F_\mt{E}(R)} \int_{\partial V} \overline{\mathcal{A}}_0^2 ~ .
    \notag
\end{align}
As in the scalar-field case, we subtract the terms that depend solely on the unperturbed Schwarzschild metric and the fixed background geometry, since they contain no information about the black hole’s response to the external electrostatic potential \eqref{eq:Maxwell_bckrnd}. As a result, we obtain
\begin{equation}
    S_{\textrm{EFT}}=S_\mt{Sch} +  { 1\over 2 }  {F'_\mt{E}(R)\over F_\mt{E}(R)}
     \int_{\partial V} \overline{\mathcal{A}}_0^2 ~.
\end{equation}
Using \eqref{use_iden1} and \eqref{use_iden2} to evaluate the spherical integral, and expanding the result for \(R\gg r_s\) while retaining only the $R$-independent term yields
\begin{align}
    S_{\textrm{EFT}}&=S_\mt{Sch}+ {\lambda_l\over 2l!}\int dt \,
    \partial_{J_1}\ldots\partial_{J_l}\overline{\mathcal{A}}_0\partial^{J_1}\ldots\partial^{J_l}\overline{\mathcal{A}}_0 ~, 
    \nonumber\\
    \lambda_l 
    &=\frac{\Omega_{d-1+2 l}}{(2\pi)^l}\frac{\left(d-3\right)}{2\left(16\right)^{\hat{l}}}\frac{\Gamma\big(\hat{l}\big)\Gamma\big(2+\hat{l}\big)}{\Gamma^{2}\big(\frac{1}{2}+\hat{l}\big)}\tan(\pi\hat{l}\,) \, r_{s}^{2 l+d-3}.
    \label{eq:Vector_eff_action_&_Love_nom.}
\end{align}
Thus, we recover the electrostatic sector of the black hole effective action, with the Love numbers in full agreement with previous calculations in the literature \cite{Hui:2020xxx}.

\subsection{Polarization Thermodynamics}

In this subsection, we examine the role of electrostatic Love numbers \eqref{eq:Vector_eff_action_&_Love_nom.} in black hole thermodynamics, building on the results derived in subsections \ref{sec:GRstresstensor}, \ref{sec:exact_gravitational_action}, and \ref{sec:smarr}. For simplicity, we restrict attention to the dipole polarization encoded in $\lambda_1$.

The analogue of \eqref{eq:Spp_scalar} is given by\footnote{Electrostatic Love numbers \eqref{eq:Vector_eff_action_&_Love_nom.} are associated with quadratic combinations of covariant derivatives of the electric component of the Maxwell field strength tensor \eqref{eq:E_mu def}. In the static limit, one has $$E_i=F_{i0}{dt\over d\tau}=e^{-\phi}\partial_i\mathcal{A}_0~, \quad E_0=0~.$$ The factor $e^{-\phi}$ is therefore responsible for the difference between the point-particle actions in the scalar and electrostatic cases.}
\begin{multline}
S_\text{p.p.}=-m\int dt~\phi+
\sum_{l=1}^\infty\frac{\lambda_{l}}{l!} \int dt~ \partial_{I_1}\ldots \partial_{I_l}\mathcal{A}_0 ~ \partial^{I_1}\ldots\partial^{I_l}\overline{\mathcal{A}_{0}}
\\
+\sum_{l=1}^\infty\frac{\lambda_{l}}{2 l!}\int dt
\Big(-l\sigma^{I_1J_1} 
+ (2\hat l-1)\phi\,\delta^{I_1 J_1}\Big)
\partial_{I_1}\partial_{I_2}\ldots\partial_{I_l}\overline{\mathcal{A}_{0}} ~ \partial_{J_1}\partial^{I_2}\ldots\partial^{I_l}\overline{\mathcal{A}_{0}} + \ldots ~,
\label{eq:Spp_vector}
\end{multline}
where we decompose the electrostatic potential into the background \eqref{eq:Maxwell_bckrnd} and a small fluctuation induced by the black hole, $\mathcal{A}_{0} \to \overline{\mathcal{A}_{0}} + \mathcal{A}_{0}$, and retain only terms linear in $\mathcal{A}_0$, $\sigma_{IJ}$, and $\phi$. The fields are weak in the regime of validity of the effective field theory, and the matter action \eqref{eq:Maxwell_action} can therefore be expanded as
\begin{equation}
     S_\mt{matter}[\mathcal{A}_{\mu}] = -\frac{1}{4}\int_V \sqrt{g}F_{\mu\nu}F^{\mu\nu} 
     =\frac{1}{2}\int_V\left(\delta^{IJ}-2\phi\delta^{IJ}-\sigma^{IJ}+\frac{1}{2}\sigma_{K}^{K}\delta^{IJ}\right)\partial_{J}A_{0}\partial_{I}A_{0} + \ldots \quad.
     \label{eq:Maxwell_mtractn}
\end{equation}
Consequently, the propagator of \(\mathcal{A}_0\) is given by
\begin{equation}
    \langle\mathcal{A}_0(x)\mathcal{A}_0(0)\rangle = \frac{1}{2\pi\Omega_{d-3}|x|^{d-3}}~.
    \label{eq:propA0}
\end{equation}

As shown in subsection \ref{sec:GRstresstensor}, the first term in \eqref{eq:Spp_vector} accounts for the energy of the unperturbed Schwarzschild black hole, given in \eqref{eq:USch}. The background configuration \eqref{eq:Maxwell_bckrnd} does not produce linear corrections to this energy, i.e., $U^{(1)} = 0$, because the energy–momentum tensor sourcing the gravitational field is quadratic in the Maxwell field.

\begin{figure}[t!]
\centering \noindent
\includegraphics[width=15cm]{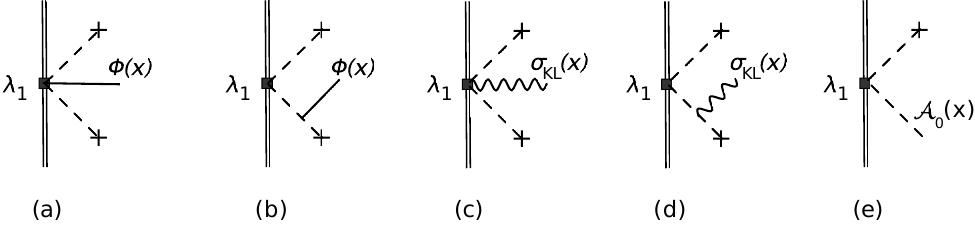}
\caption[]{Feynman diagrams contributing to the asymptotic values of the gravitational fields $\phi$, $\sigma_{KL}$, and $\mathcal{A}_0$ in the EFT approach. The black hole worldline is represented by a double line. Dashed lines denote $\mathcal{A}_0$ propagators, while a dashed line with a cross represents the background field $\overline{\mathcal{A}_0}$. The heavy boxes in panels (a) and (c) correspond to vertices associated with the two terms in the second line of \eqref{eq:Spp_vector}, whereas those in panels (b), (d), and (e) correspond to the second term in \eqref{eq:Spp_vector}.}
\label{Fig:BHenergyM}
\end{figure}

The leading correction to $U_{\text{Sch}}$ therefore arises at quadratic order in $\overline{\mathcal{A}_0}$. The $R$-independent contributions are captured by the diagrams shown in Fig.~\ref{Fig:BHenergyM}(a)-(d). Using \eqref{eq:propPhi}, \eqref{eq:propSigma}, \eqref{eq:Spp_vector}, \eqref{eq:Maxwell_mtractn}, and \eqref{eq:propA0}, one finds
\begin{align}
  &   \text{Fig.}\ref{Fig:BHenergyM}\text{(a)}=-\frac{2\left(d-5\right)}{\left(d-2\right)\Omega_{d-3}} ~ \frac{\lambda_{1}\mathcal{E}_K\mathcal{E}^K}{|x|^{d-3}} ~,
  \nonumber\\
  &   \text{Fig.}\ref{Fig:BHenergyM}\text{(b)}=\frac{8\pi\left(d-3\right)\lambda_{1}}{\left(d-2\right)\Omega_{d-1}|x|^{d-1}}\left(\frac{\mathcal{E}_{K}\mathcal{E}^{K}}{d-3}x^{2}-\left(\mathcal{E}_{K}x^{K}\right){}^{2}\right)~,
\nonumber\\
  &  \text{Fig.}\ref{Fig:BHenergyM}\text{(c)}=\frac{8\lambda_{1}}{\Omega_{d-3}|x|^{d-3}}\left(\frac{\mathcal{E}_K\mathcal{E}^K}{d-3}\delta_{KL}-\mathcal{E}_{K}\mathcal{E}_{L}\right)~,
\\
    &\text{Fig.}\ref{Fig:BHenergyM}\text{(d)} = \frac{8\lambda_{1}}{\Omega_{d-3}|x|^{d-1}}\left(\mathcal{E}_{K}\mathcal{E}_{L}\,x^{2}-\left(d-3\right)\mathcal{E}_{I}x^{I}\,\mathcal{E}_{(K}x_{L)}\right) ~,
\nonumber\\
&\text{Fig.}\ref{Fig:BHenergyM}\text{(e)}=-\frac{\lambda_1\overline{\mathcal{A}_{0}}}{\Omega_{d-1}|x|^{d-1}}~.
    \nonumber
\end{align}
As in the scalar case, for most diagrams no computation beyond the Feynman rules is required. Figures \ref{Fig:BHenergyM}(b) and \ref{Fig:BHenergyM}(d) are the only ones that involve a single integral over a bulk point. The relevant integral is evaluated in Appendix \ref{apx:integrals}. As expected, the above expressions for $\sigma_{KL}$ satisfy the harmonic gauge condition \eqref{eq:GF_action}, namely, $\frac{1}{2}\partial_{I}\sigma_{\,J}^{J}=\partial_{J}\sigma_{I}^{\,J}$.

Substituting into \eqref{eq:U_with_NRG} and including the background energy computed in Appendix \ref{appx:bckgrnd_energy} then yields the second-order correction to $U_{\text{Sch}}$
\begin{equation}
    U^{(2)}_\infty=\frac{\Omega_{d-1}}{2(d-1)}R^{d-1}\mathcal{E}_{J}\mathcal{E}^{J}-
    \frac{1}{2}\frac{d-3}{d-1}\lambda_{1}\mathcal{E}_{J}\mathcal{E}^{J}+\ldots~,
\end{equation}
where the ellipsis denotes contributions that depend on both $R$ and the black hole mass $m$, and the subscript $\infty$ indicates that this energy corresponds to electrostatic potential that approaches $\overline{\mathcal{A}_0}$ at spatial infinity, rather than at finite radius $r = R$ as in \eqref{eq:vector_full}.

To adapt this result to the boundary condition imposed in our setup – where all fields match the background at finite radius $r = R$ instead of at infinity – it is sufficient to rescale the perturbation amplitude by a constant factor,
\begin{equation}
\mathcal{E}_J \to {\mathcal{E}_J\over F_\mt{E}(R)}~, \quad 
F_\mt{E}(R) = 1 -  \frac{\lambda_{1}}{\Omega_{d-1}R^{d-1}} 
+\ldots \quad .
\end{equation}
The expression for $F_\mt{E}(R)$ given above follows from the diagram in Fig.~\ref{Fig:BHenergyM}\text{(e)}, and it is consistent with the asymptotic expansion of \eqref{eq:FE}. Additional contributions depend on the black hole mass $m$, but they are not relevant for our purposes. As a result, we obtain
\begin{equation}
    U^{(2)}=\frac{\lambda_{1}}{2}\mathcal{E}_{I}\mathcal{E}^{I}  ~,
\end{equation}
where the $R$-dependent terms – including those that depend on $m$ – have been subtracted (see footnote \ref{foot:scheme-dependent}).

On the other hand, using \eqref{eq:action_perturbation_def}, \eqref{eq:free_energy_is_action}, and \eqref{eq:Vector_eff_action_&_Love_nom.}, the free energy associated with the polarization of the black hole is given by

\begin{equation}
F^{(2)} =  -\frac{\lambda_1}{2}\mathcal{E}_{J}\mathcal{E}^{J} .
\end{equation}
The second-order correction to $TS$ then reads
\begin{equation}
    \delta^{(2)}(TS)=U^{(2)}-F^{(2)} =
    \lambda_{1}\mathcal{E}_{I}\mathcal{E}^{I} \quad .
\end{equation}

Substituting these expressions into \eqref{eq:QP_F_U}, we obtain
\begin{equation}
    \mathcal{M}_{I}\mathcal{P}^{I} = -\frac{d-1}{2}\lambda_{1}\mathcal{E}_{I}\mathcal{E}^{I}~.
\end{equation}
Identifying $\mathcal{M}_J$ with the external control parameter $-\mathcal{E}_J$ then yields the desired polarization vector,
\begin{equation}
    \mathcal{P}^{I} = \frac{d-1}{2}\lambda_{1}\mathcal{E}^{I} ~.
\end{equation}

\section{Gravito-magnetic background}
\label{sec:gravito-magnetic}

In the previous sections, we examined static perturbations of the Schwarzschild black hole induced by matter fields. These perturbations arose from external sources and were described by the scalar and electromagnetic fields. In this section, we turn our attention to stationary perturbations of purely gravitational origin. To this end, we set the matter action to zero, $S_\text{matter}=0$, thereby removing external matter sources from the system. We then introduce a gravito-magnetic background, which captures stationary but non-static gravitational perturbations of the Schwarzschild geometry
\begin{equation}
\overline{A}_{I}:=C_{I J_{1} \cdots J_{l}}x^{J_{1}}...x^{J_{l}} ~, \quad \overline{\phi}=0~, \quad \overline\gamma_{IJ}=\delta_{IJ} ~,
\label{eq:gravito-magnetic-bckgnd}
\end{equation}
where $C_{I J_{1} \cdots J_{l}}$ is symmetric in all the last $l$ indices, totally traceless, and obeys the condition
\begin{equation}
C_{I J_{1} \cdots J_{l}}+C_{J_{l}I J_1...J_{l-1}}+...+C_{J_{1}...J_{l}I}=0~.
\label{eq:cyclic}
\end{equation}
This background solves the linearized Einstein equations \eqref{eq:EOM_NRG} and satisfies the harmonic gauge condition,\footnote{The linearization of the Einstein equations is justified in a sufficiently small neighborhood around the origin where $\overline{A}_{I}\ll 1$.} 
\begin{equation}
\partial_{I}\overline{F}^{IJ}=0~, \quad \overline A_r=0 ~, \quad \nabla^{\mathbb{S}^{d-2}}_i \, \overline A^i =0 ~.
\label{backGauge}
\end{equation}
Indeed, 
\begin{equation}
\overline{A}_{r}={\partial x^I\over \partial r} \, \overline{A}_I={x^I\over r} \, C_{I J_{1},..,J_{l}} x^{J_{1}}...x^{J_{l}}=0 ~,
\end{equation}
where the last equality is due to \eqref{eq:cyclic}. Hence, $\nabla^{\mathbb{S}^{d-2}}_i \, \overline A^i =0$ follows from $\partial_I \overline A^I=0$ written in spherical coordinates with $\overline A_r=0$. 

The nonzero components of the linear perturbation $g^\mt{(1)}_{\mu\nu}$ take the form \cite{Hadad:2024lsf},
\begin{equation}
 g^\mt{(1)}_{it}(r,\Omega)=-2f(r)\,A^\mt{full}_i(r,\Omega)~, \quad A_I^\mt{full}(r,\Omega)= {\overline A_I(r,\Omega) \over f(R) F_\mt{GM}(R)} F_\mt{GM}(r) ~,
\label{eq:Afull}
\end{equation}
where a radial profile that is smooth at the horizon is given by
\begin{equation}
 F_\mt{GM}(r)=\hat{l}_+\big(\hat{l}_+ +1\big)B\big(1+\hat{l}_- \,,\hat{l}_+ \,\big)
 \;{}_2F_1\Big(~1 - \hat{l}_-~,~ -\hat{l}_+~,~2~;1-\Big({r_s\over r}\Big)^{d-3}\Big)
 ~, \quad \hat{l}_\pm={l\pm 1\over d-3} ~.
\end{equation}
Note that the amplitude of the perturbation $A_I^\mt{full}$ is normalized to satisfy the boundary condition
\begin{equation}
 g^\mt{(1)}_{It}(r,\Omega)|_{r=R} = -2 \overline A_I(R,\Omega) ~,
 \label{eq:boundary_metric_GM}
\end{equation}
which ensures that the linear perturbation matches the gravito-magnetic background \eqref{eq:gravito-magnetic-bckgnd} at $r=R$. 

Because the linear perturbation is off-diagonal, whereas the Schwarzschild metric \eqref{eq:Schw}, the unperturbed induced metric, and the extrinsic curvature of $\partial V$ \eqref{eq:K_Sch} are diagonal, the linear correction \eqref{eq:d1S_full} vanishes, while the quadratic correction \eqref{eq:d2S_full} simplifies, yielding\footnote{Recall that indices are raised and lowered with $g_{\mu\nu}^S$.}
\begin{align}
S_\mt{full}&=S_\mt{Sch} + {1\over 16\pi} 
\int_{\partial V} \sqrt{|h|} \Big(\mathcal{K}_S\, h^{\mu\nu} - \mathcal{K}_S^{\mu\nu} \Big) g^\mt{(2)}_{\mu\nu} 
\nonumber \\
&- {1\over 16\pi}  \int_{\partial V} \sqrt{|h|}
\Big(   \mathcal{K}_S\, g^{\mt{(1)}it} +  g_S^{i\alpha}g_S^{t\beta}\delta^{(1)} \mathcal{K}_{\alpha\beta} - \mathcal{K}^{ij}_{S} \, g^{\mt{(1)}\,t}_{~j}-\mathcal{K}_S^{tt}g_S^{ij} g^\mt{(1)}_{tj}\Big) g^\mt{(1)}_{it} + \ldots ~.
\end{align}
The ellipsis encodes terms that are cubic or higher in the weak perturbation \eqref{eq:gravito-magnetic-bckgnd}. Based on Appendix \ref{app:curvature_variations} and the structure of the unperturbed Schwarzschild metric \eqref{eq:Schw}, we have
\begin{align}
 & g_S^{i\alpha}g_S^{t\beta}\delta^{(1)} \mathcal{K}_{\alpha\beta} =  {\Omega^{ij}\over R^2\sqrt{f(R)}} \, {\partial\over\partial r}\big(fA^\mt{full}_j\big)\Big|_{r=R} ~.
\end{align}
Hence,
\begin{align}
&S_\mt{full}=S_\mt{Sch}+{1\over 8\pi}   R^{d-4}f(R)\int dt\int d\Omega~
 \Omega^{ij} \overline A_i {\partial \overline A^\mt{full}_j \over \partial r}\Big|_{r=R}
 \\
 &+ {1\over 16\pi} 
\int_{\partial V} \sqrt{|h|} \Big(\mathcal{K}_S h^{\mu\nu} - \mathcal{K}_S^{\mu\nu} \Big) g^\mt{(2)}_{\mu\nu} 
+ {1\over 8\pi}   R^{d-5}\Big(d+1+{d-3\over f(R)}\Big)\int dt\int d\Omega
 \overline A_i \overline A_j \Omega^{ij}   + \ldots
\nonumber
\end{align}
Substituting \eqref{eq:Afull}, we can rewrite it as follows
\begin{align}
&S_\mt{full}=S_\mt{Sch} + {1\over 8\pi}   R^{d-4}{F'_\mt{GM}(R)\over F_\mt{GM}(R)}
 \int dt\int d\Omega ~\overline A_i \overline A_j \Omega^{ij}  
 \label{SfullPert}\\
 &+ {1\over 16\pi} 
\int_{\partial V} \sqrt{|h|} \Big(\mathcal{K}_S h^{\mu\nu} - \mathcal{K}_S^{\mu\nu} \Big) g^\mt{(2)}_{\mu\nu} 
+ {1\over 8\pi}R^{d-5}\Big(d+l+2+{d-3\over f(R)}\Big)\int dt\int d\Omega
 \overline A_i \overline A_j \Omega^{ij}
    + \ldots
\nonumber
\end{align}

The expressions in the second line are independent of $F_\mt{GM}$. They are fully determined by the external background \eqref{eq:gravito-magnetic-bckgnd} and the boundary conditions imposed on $\partial V$. These terms correspond to the self-energy of the perturbation \eqref{eq:gravito-magnetic-bckgnd} and to the interaction energy between the unperturbed Schwarzschild black hole and the external background, without accounting for the black hole’s response. Therefore, both expressions can be treated as reference geometry and subtracted without affecting the gravito-magnetic Love numbers, allowing us to retain only the terms in the first line, which depend on the response function $F_\mt{GM}$.

The spherical integral can be evaluated by noting that, due to $\overline A_r=0$, we have
\begin{equation}
    \overline A_i \overline A_j \Omega^{ij} = R^2 \delta^{IJ} \overline A_I \overline A_J ~.
\end{equation}
Hence,
\begin{equation}
   \int d\Omega \, \Omega^{ij} \, \overline A_i\overline A_j  = 
   R^2 C_{I J_{1} \cdots J_{l}} 
   C^I_{~I_{1} \cdots I_{l}}
   \int d\Omega ~ x^{J_{1}}...x^{J_{l}} x^{I_{1}}...x^{I_{l}}  ~.
\end{equation}
Using \eqref{use_iden1} and \eqref{use_iden2}, we obtain
\begin{equation}
 \int d\Omega \, \overline A_i\overline A_j \Omega^{ij} = {\Omega_{d-1+2l}\over (2\pi)^l} ~ l! ~ R^{2l+2} \, 
 C_{I J_{1} \cdots J_{l}} C^{I J_{1} \cdots J_{l}} ~.
\end{equation}
Substituting into \eqref{SfullPert} yields
\begin{align}
S_\mt{full}= S_\mt{Sch} +
  {\Omega_{d-1+2l}\over 8\pi(2\pi)^l}\,l!
   R^{2l+d-2}{F'_\mt{GM}(R)\over F_\mt{GM}(R)}\int dt \, C_{I J_{1} \cdots J_{l}} C^{I J_{1} \cdots J_{l}}+ \ldots
\end{align}
Furthermore, in the EFT domain, where $R\gg r_s$, one has
\begin{equation}
 F_\mt{GM}(R)=1 + \,{B(2+\hat{l}_+ \,,\hat{l}_- \,) \over B(-\hat{l}_+\,, \, -\hat{l}_-)}  \Big({r_s\over R}\Big)^{2 l+d-3} +\ldots ~.
  \label{sol_for_A}
\end{equation}
Hence, using \eqref{eq:action_perturbation_def}, we obtain 
\begin{align}
 S_{\textrm{EFT}}=S_\mt{Sch}  - {\Omega_{d-3+2l}\over 4(2\pi)^l} ~ l!
 \,{B(2+\hat{l}_+ \,,\hat{l}_- \,) \over B(-\hat{l}_+\,, \, -\hat{l}_-)} r_s^{2l+d-3}\int dt \, C_{I J_{1} \cdots J_{l}} C^{I J_{1} \cdots J_{l}}+ \ldots ~.
 \label{LoveFull}
\end{align}
where, as before, the $R$-dependent terms have been suppressed (see footnote \ref{foot:scheme-dependent}). Using \eqref{eq:gravito-magnetic-bckgnd}, this expression can be recast in the form \eqref{eq:PP_action_weak_fields} with the gravito-magnetic Love number $C_l^B$ matching the established result in the literature \eqref{eq:LoveNum}.

\subsection{Thermal corrections}

Let us compute the polarization of the black hole in the presence of a weakly curved gravito-magnetic background \eqref{eq:gravito-magnetic-bckgnd}. In contrast to the scalar and electromagnetic perturbations studied in the previous sections, the simplest nontrivial case for a purely gravitational perturbation corresponds to $l=2$, which we analyze here. The $l=1$ mode is pure gauge, as the Riemann tensor for such a background vanishes identically.

The expression \eqref{eq:U_with_NRG} for the total energy is quadratic in $A_I$; therefore, there is no linear correction to the unperturbed energy \eqref{eq:USch} of the Schwarzschild black hole. As before, we evaluate the leading second order correction. To this end, we decompose the total metric into the background \eqref{eq:gravito-magnetic-bckgnd} and a small response of the black hole, $A_I\to \overline{A}_I+A_I$, retaining terms that are at most quadratic in $\overline{A}_I$. In particular, the black hole effective action \eqref{eq:PP_action} takes the form
\begin{multline}
S_\text{p.p.}=-m \int dt \, \phi+
\frac{C_{2}^B}{2} \int dt \, \partial_{K}F_{IJ} ~ \partial^K
\overline{F}^{IJ}
\label{eq:Spp_GM}
\\
+{C_{2}^B\over 4} \int dt \Big(\frac{3(d-1)}{d-3} \partial_{K}\overline{F}_{IJ} ~ \partial^K
\overline{F}^{IJ}\, \phi - \sigma^{KL}\partial_{K}\overline{F}_{IJ} ~ \partial_L
\overline{F}^{IJ} - 2\,\sigma^{JL}\,
\partial_{K}\overline{F}_{IJ} ~ \partial^K
\overline{F}^{I}_{~L} \Big)  
+ \ldots ~,
\end{multline}
where only the linear terms in the response are shown, while higher-order contributions are indicated by ellipses.\footnote{Worldline terms that are nonlinear in the response do not contribute to the asymptotic value of the perturbation. There are also cubic terms – quadratic in $\overline{A}_I$ and linear in $A_I$ – associated with Love numbers multiplying operators cubic in the curvature; however, their contribution to thermodynamics decays too rapidly at infinity to affect the analysis at second order in $\overline{A}_I$.} Here, and throughout the remainder of this section, Latin indices are raised and lowered using the Kronecker delta $\delta_{IJ}$.

\begin{figure}[t!]
\centering \noindent
\includegraphics[width=15cm]{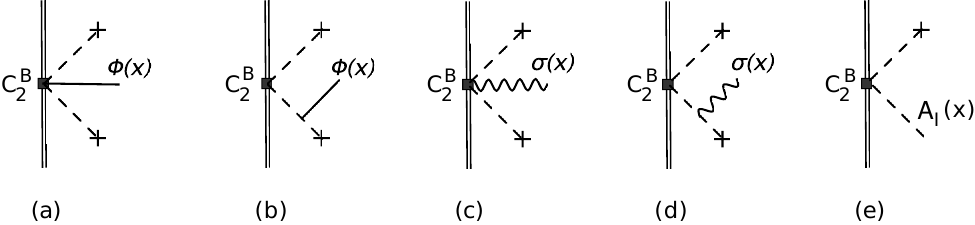}
\caption[]{Feynman diagrams contributing to the asymptotic values of the gravitational fields $\phi$, $\sigma=\delta^{KL}\sigma_{KL}$, and $A_I$ in the EFT approach. The black hole worldline is represented by a double line. Dashed lines denote $A_I$ propagators \eqref{eq:propA}, while a dashed line with a cross represents the background field \eqref{eq:gravito-magnetic-bckgnd}. The heavy boxes in panels (a) and (c) correspond to vertices associated with the second line of \eqref{eq:Spp_GM}, whereas those in panels (b), (d), and (e) correspond to the last term in the first line of \eqref{eq:Spp_GM}.}
\label{Fig:BHenergyGM}
\end{figure}

The $R$-independent contributions to $U^{(2)}$ are captured by the diagrams shown in Fig.~\ref{Fig:BHenergyGM}(a)-(d), where the cubic vertices in the bulk are obtained by expanding the Einstein-Hilbert action \eqref{eq:EH_action} in the weak fields,
\begin{equation}
    S_\mt{EH} \supset {d-2\over 4\pi(d-3)} \int_V \phi\, F_{IJ}\overline{F}^{IJ}+{1\over 8\pi}\int_V\Big(\sigma \,\partial_IA_J\overline{F}^{IJ} - 2\sigma^{IJ} F_{IK}\overline{F}_J^{~~K} \Big) ~ ,
\end{equation}
where $\sigma=\delta^{KL}\sigma_{KL}$.Using \eqref{eq:propPhi}, \eqref{eq:propA},  and \eqref{eq:propSigma}, one finds\footnote{Note that $\partial_K \overline{F}_{IJ}\,\partial^K\overline{F}^{IJ}=12 \, C_{IJK}C^{IJK}$.}
\begin{align}
  &   \text{Fig.}\ref{Fig:BHenergyGM}\text{(a)}={36(d-1)\over (d-2)\Omega_{d-3}} {C_2^B\,C_{IJK}C^{IJK}\over |x|^{d-3}}
  ~,
  \nonumber\\
  &   \text{Fig.}\ref{Fig:BHenergyGM}\text{(b)}=
  {18(d-3)C_2^B\over \Omega_{d-3}}
  \Big(-{2\over d-3}{C_{IJK}C^{IJK}\over |x|^{d-3}} +(d-1){C_{ILM} C^{IKN}x^L x^M x_K x_N \over |x|^{d+1}}
  \nonumber\\
 &\quad\quad\quad\quad\quad\quad\quad\quad\quad\quad\quad~
 +{4\over 3}{C_{IJK}C^{IJL} x^K x_L\over |x|^{d-1}}+{8\over 3} {C_{IJK}C^{JIL}x^Kx_L\over |x|^{d-1}}\Big)~,
\nonumber\\
  &  \text{Fig.}\ref{Fig:BHenergyGM}\text{(c)}={288\over (d-3)\Omega_{d-3}} {C_2^B\,C_{IJK}C^{IJK}\over |x|^{d-3}}
  ~,
\\
    &\text{Fig.}\ref{Fig:BHenergyGM}\text{(d)} =  
    {48(d-5)\over \Omega_{d-3}}{C_2^B\over |x|^{d-3}} \Big({3\over 2(d-3)}C_{IJK}C^{IJK} 
    -{3(d-1)\over 4}{C_{ILM} C^{IKN}x^L x^M x_K x_N \over |x|^{4}}
    \nonumber\\
&\quad\quad\quad\quad\quad\quad\quad\quad\quad\quad\quad\quad\quad
    -{C_{IJK}C^{IJL} x^K x_L\over |x|^{2}}-2{C_{IJK}C^{JIL}x^Kx_L\over |x|^{2}}\Big)
    ~,
\nonumber\\
&\text{Fig.}\ref{Fig:BHenergyGM}\text{(e)}=-{24\pi^2\over\Omega_{d+1}} C_2^B {\overline{A}_I\over |x|^{d+1}}
    ~.
    \nonumber
\end{align}
As in the previous sections, most diagrams require no computation beyond the Feynman rules. Only Figures \ref{Fig:BHenergyGM}(b) and \ref{Fig:BHenergyGM}(d) involve a single bulk integral, evaluated in Appendix \ref{apx:integrals}.

Substituting the harmonic gauge condition $\partial_{J}\sigma_{I}^{\,J}=\frac{1}{2}\partial_{I}\sigma$ into \eqref{eq:U_with_NRG} and evaluating separately the contribution of each diagram yields\footnote{Note that individual diagrams do not necessarily satisfy the harmonic gauge condition; however, their sum does, and it is this sum that ultimately enters the total energy calculation.}
\begin{align}
    & U^{(2)}_{\text{Fig.}\ref{Fig:BHenergyGM}\text{(a)}}= -9 {d-1\over d-3}C_2^BC_{IJK}C^{IJK}  ~, \quad U^{(2)}_{\text{Fig.}\ref{Fig:BHenergyGM}\text{(b)}}={36(d-2)\over (d+1)(d-3)}C_2^BC_{IJK}C^{IJK}~, 
    \nonumber \\
    & U^{(2)}_{\text{Fig.}\ref{Fig:BHenergyGM}\text{(c)}}={18\over d-3}C_2^B C_{IJK}C^{IJK}~, \quad
    U^{(2)}_{\text{Fig.}\ref{Fig:BHenergyGM}\text{(d)}}={18(d-5)\over (d+1)(d-3)} C_2^BC_{IJK}C^{IJK}~,
     \nonumber \\
     & U^{(2)}_{\text{Fig.}\ref{Fig:BHenergyGM}\text{(e)}}={12(d-4)\over d+1}C_2^B C_{IJK}C^{IJK}~.
\end{align}

Upon combining all contributions and adding the background energy computed in Appendix \ref{appx:bckgrnd_energy}, one obtains 
\begin{equation}
    U^{(2)}_\infty=\frac{\Omega_{d-1}}{4\pi(d-1)(d+1)}R^{d+1}C_{IJK}C^{IJK}+
    {3(d-1)\over (d+1)} C_2^B C_{IJK}C^{IJK}+\ldots~.
\end{equation}
As before, ellipsis denotes contributions that depend on both $R$ and the black hole mass $m$. The subscript $\infty$ indicates that this energy correction corresponds to a Schwarzschild black hole in the presence of a weak gravito-magnetic background, for which the boundary condition \eqref{eq:boundary_metric_GM} is imposed at spatial infinity rather than at a finite radius $r = R$.

To impose \eqref{eq:boundary_metric_GM}, it is sufficient to rescale the perturbation amplitude by a constant factor,
\begin{equation}
C_{IJK} \to {C_{IJK}\over F_\mt{GM}(R)}~, \quad 
F_\mt{GM}(R) = 1 -{24\pi^2\over\Omega_{d+1}}  {C_2^B\over R^{d+1}} 
+\ldots \quad .
\end{equation}
The expression for $F_\mt{GM}(R)$ given above follows from the diagram in Fig.~\ref{Fig:BHenergyGM}\text{(e)}. Additional contributions depend on the black hole mass $m$, but they are not relevant for our purposes. As a result, we obtain
\begin{equation}
    U^{(2)}=3\,C_2^B C_{IJK}C^{IJK}  ~,
\end{equation}
where the $R$-dependent terms – including those that depend on $m$ – have been subtracted (see footnote \ref{foot:scheme-dependent}).

Using \eqref{eq:PP_action_weak_fields}, \eqref{eq:action_perturbation_def}, and \eqref{eq:free_energy_is_action}, the free energy associated with the polarization of the black hole is given by
\begin{equation}
F^{(2)} = - 3\,C_2^B C_{IJK}C^{IJK} .
\end{equation}
The second-order correction to $TS$ then reads
\begin{equation}
    \delta^{(2)}(TS)=U^{(2)}-F^{(2)} =6\,C_2^B C_{IJK}C^{IJK} \quad .
\end{equation}
Substituting these expressions into \eqref{eq:QP_F_U}, we obtain
\begin{equation}
\mathcal{M} \mathcal{P}= - {3(d-1)\over 2}\,C_2^B C_{IJK}C^{IJK} .
\end{equation}
Identifying $\mathcal{M}$ with the gauge invariant external control parameter $-\partial_K\overline{F}_{IJ}$ then yields the desired polarization vector,
\begin{equation}
    \mathcal{P}^{KIJ}= {d-1\over 2}C_2^B \, C^{[JI]K}\quad .
\end{equation}

\section{Gravito-electric background}
\label{sec:gravito-electric}

The weakly curved gravito-electric background is defined by 
\begin{align}
\label{eq:gravito_electric_background_def}
\overline{\phi}:=Q_{J_{1}...J_{l}}x^{J_{1}}...x^{J_{l}},\,\,\,\overline{A}_{I}=0,\,\,\,\,\overline{\gamma}_{IJ}=\delta_{IJ}
\end{align}
where \(Q_{J_{1}...J_{l}}\) is a totally symmetric, traceless tensor. This configuration solves the linearized Einstein equations and carries no energy. At second order in the amplitude \(Q_{J_{1}...J_{l}}\), the equations of motion \eqref{eq:EOM_NRG} induce corrections \(\overline{\sigma}_{IJ}\) to the spatial metric of the background. The vector field, however, remains \(\overline{A}_I=0\) to all orders in \(Q_{J_{1}...J_{l}}\) as required by the staticity of the perturbation.

In the previous sections, we derived the effective action by explicitly evaluating the perturbative corrections to the full gravitational action on shell, thereby directly obtaining the Love number associated with each type of perturbation without resorting to the standard matching procedure. The fields associated with these perturbative backgrounds decouple from one another at linear order, which considerably simplifies their treatment. In the gravito-electric case, however, this procedure becomes technically more involved, as the equations of motion \eqref{eq:EOM_NRG} mix $\phi$ and $\sigma_{IJ}$ at linear order. The master field approach \cite{Regge:1957td,Zerilli:1970se,Kodama:2003jz,Ishibashi:2003ap} allows one to resolve this mixing by reducing the Einstein equations to a single ordinary differential equation of hypergeometric type. In this approach, the linear perturbation $g^\mt{(1)}_{tt}$ takes the form \cite{Kodama:2003jz}
\begin{equation}
   g^\mt{(1)}_{tt}(r,\Omega)=2f(r) \phi^\mt{full}(r,\Omega)~, \quad \phi^\mt{full}(r,\Omega)={\overline{\phi}(R,\Omega)\over f(R) F_\mt{GE}(R)} F_\mt{GE}(r)
   \label{eq:GEscalar}
\end{equation}
where a radial profile that is smooth at the horizon is given by\footnote{In general, this relation contains additional gauge-dependent terms, denoted by $f_r$ and $H_T$ in \cite{Kodama:2003jz}. For simplicity, we set them to zero by an appropriate choice of gauge. \label{foot-gauge}}
\begin{eqnarray}
    F_\mt{GE}(r)&=&{2(d-3)\over d-2}{f(r)Y'(r)\over r^{d-4}f'(r)}  + {d-4\over d-2}\,{Y(r)\over r^{d-4}}~, 
    \non
    Y(r)&=& \Big({r_s\over r}\Big)^{1-l}\;{}_2F_1\Big(-\hat{l}~,~ 1-\hat{l}~,~2~;1-\Big({r_s\over r}\Big)^{d-3}\Big) ~.
    \label{eq:GEmaster}
\end{eqnarray}
The amplitude $\phi^\mt{full}$ is normalized such that
\begin{equation}
    g^\mt{(1)}_{tt}(r,\Omega)|_{r=R}=2\overline\phi~.
    \label{eq:GS_bndry_cond}
\end{equation}
As argued in Appendix~\ref{app:full_action_GE_consistency}, this profile is sufficient to evaluate the action on shell up to second order in the gravito-electric perturbation. Using \eqref{eq:d2S_relevant_GE_simple} and \eqref{eq:GEscalar}, we obtain
\begin{align}
\delta^{\left(2\right)}S&=-\frac{1}{8\pi}\frac{\left(d-2\right)}{\left(d-3\right)}fR^{d-2}\int_{\partial V}d\Omega_{d-2}\phi\partial_{r}\phi \non &-\frac{1}{32\pi}f^{\frac{d-5}{d-3}}R^{d-6}\int_{\partial V}d\Omega_{d-2}\left(\Omega^{ik}\Omega^{jl}-\Omega^{ij}\Omega^{kl}\right)\sigma_{kl}\partial_{r}\sigma_{ij}=C_{l}^{E}
Q_{J_{1}...J_{l}}Q^{J_{1}...J_{l}} +\ldots ~,
\label{eq:GSfullS}
\end{align}
where $C_l^E$ agrees with \eqref{eq:LoveNum}. Note that the term quadratic in $\sigma_{ij}$ does not contribute, since $\sigma_{ij}$ decays faster than $\phi$ at infinity. The final $R$-independent constant arises entirely from the term quadratic in $\phi$. To check this explicitly, one can use the expression for $\sigma_{ij}$ in terms of the master field  \cite{Kodama:2003jz}
\begin{equation}
    \sigma_{ij} =  - {2\overline{\phi}(R,\Omega)\over f(R) F_\mt{GE}(R)} {Y(r) f^{1\over d-3}\over (d-3) r^{d-6}} \, \Omega_{ij} ~,
\end{equation}
where the gauge choice (see footnote \ref{foot-gauge}) and normalization constant are consistent with \eqref{eq:GEscalar}.

In Appendix~\ref{app:full_action_GE_consistency}, we present an instructive consistency check. There, rather than using the master field of the full theory, we evaluate the full action on shell using the metric components in Table~\ref{tab:gravito_electric_diagrams}, obtained from the complementary EFT approach described below, for the special case $l=2$. The resulting expression agrees with \eqref{eq:GSfullS}.

To perform the EFT calculations, we impose the harmonic gauge condition \eqref{eq:GF_action}. As in the previous sections, we split the metric into the background \eqref{eq:gravito_electric_background_def} and the response of the Schwarzschild black hole, and diagrammatically evaluate its asymptotic behavior, which is sufficient to determine the corrections to various thermodynamic quantities. We restrict our attention to interaction terms involving the black hole's finite-size parameter $C_2^E$ and the background, as these encode information about deviations from the pure Schwarzschild geometry. The leading nontrivial contributions to the thermodynamic quantities are quadratic in \(Q_{IJ}\). 

Substituting \(\phi \to \overline{\phi} +  \phi\) and  \(\gamma_{IJ}\to\delta_{IJ}+\sigma_{IJ}\) and retaining only terms linear in \(\phi\) and \(\sigma_{IJ}\), the point particle action becomes 
\begin{equation}
\label{eq:pp_action_with_GE_background}
S_\mt{p.p.}\supset-m\int dt \, \phi + \frac{C_{2}^{E}}{2} \int dt \, \partial_{I}\partial_{J}\phi\,\partial^{I}\partial^{J}\overline{\phi} \notag +\frac{C_{2}^{E}}{2}\int dt\Big(\frac{d+1}{2\left(d-3\right)}\phi\,\delta^{IJ}-\sigma^{IJ}\Big)\partial_{I}\partial_{K}\overline{\phi}\partial_{J}\partial^{K}\overline{\phi}\,.
\end{equation}
The relevant cubic interaction in the bulk reads,
\begin{align}
\label{eq:EH_actions_with_A=0}
S_{\textrm{EH}}
\supset -\frac{1}{16\pi}\frac{d-2}{d-3}\int_{V} \partial_{I}\phi\partial_{J}\phi\left(\frac{\sigma}{2}\delta^{IJ}-\sigma^{IJ}\right).
\end{align}
These terms in the effective action \eqref{eq:full_EFT_action}, together with the propagators \eqref{eq:propPhi} and \eqref{eq:propSigma}, constitute the building blocks of the diagrams summarized in  Table~\ref{tab:gravito_electric_diagrams}.
Although the individual graphs contributing to \(\sigma_{IJ}\) (figures (c) and (d) in Table~\ref{tab:gravito_electric_diagrams}) do not separately satisfy the harmonic gauge condition, their sum does, as expected at each order in the perturbative expansion.

\renewcommand{\arraystretch}{2.5}
\setlength{\abovecaptionskip}{0pt}
\begin{table}[t]
\begin{tabular}{@{}c c >{\raggedright\arraybackslash}m{10cm}@{}}

\multirow{2}{1em}{(a)} & \multirow{2}{8em}{\includegraphics[width=.2\textwidth]{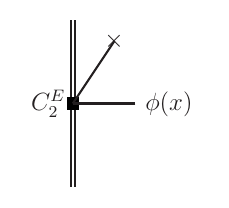}} & \raisebox{0\height}{\( = C_{2}^{E}Q^{IJ}\frac{8\pi\left(d-3\right)\left(d-1\right)}{\left(d-2\right)\Omega_{d-1}}\frac{\mathbf{\mathbf{x}}_{I}\mathbf{\mathbf{x}}_{J}}{\left|\mathbf{x}\right|^{d+1}}\)} \\
& & \raisebox{1\height}{\(\Delta U^{\left(2\right)}=\frac{4}{d+1}C_{2}^{E}Q_{IJ}Q^{IJ}\)} \\
\hdashline
\multirow{2}{1em}{(b)} & \multirow{2}{8em}{\includegraphics[width=.2\textwidth]{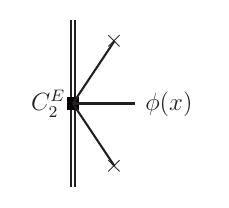}} & \raisebox{0\height}{\( = C_{2}^{E}Q_{IJ}Q^{IJ}\frac{4\left(d+1\right)}{\left(d-2\right)\Omega_{d-3}\left|x\right|^{d-3}}\)}\\
& & \raisebox{1\height}{\(\Delta U^{\left(2\right)}=-\left(\frac{d+1}{d-3}\right)C_{2}^{E}Q_{IJ}Q^{IJ}\)} \\
\hdashline
\multirow{2}{1em}{(c)} & \multirow{2}{8em}{\includegraphics[width=.2\textwidth]{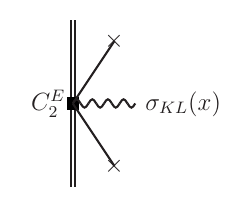}} & \raisebox{0\height}{\( = -C_{2}^{E}\frac{32}{\Omega_{d-3}\left|x\right|^{d-3}}\left(Q_{KM}Q_{L}^{\,M}-\frac{1}{d-3}\delta_{KL}Q^{IJ}Q_{IJ}\right)\)} \\
& & \raisebox{1\height}{\(\Delta U^{\left(2\right)}=8\frac{\left(d-2\right)}{\left(d-3\right)\left(d-1\right)}C_{2}^{E}Q_{IJ}Q^{IJ}\)} \\
\hdashline
\multirow{3}{1em}{(d)} & \multirow{3}{8em}{\includegraphics[width=.2\textwidth]{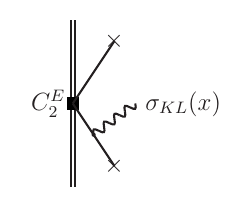}} & \raisebox{0\height}{\(
 = -\frac{8\left(d-3\right)}{\Omega_{d-3}}C_{2}^{E}\Bigg(\left(d-1\right)\frac{Q^{IJ}Q_{\,(K}^{M}\mathbf{x}_{L)}\mathbf{x}_{M}\mathbf{x}_{I}\mathbf{x}_{J}}{\left|\mathbf{x}\right|^{d+1}}\)} \\
& & \raisebox{1\height}{\(\,\,\,\,\,+\frac{2Q^{MJ}Q_{M(K}\mathbf{x}_{L)}\mathbf{x}_{J}-2Q_{(K}^{\,J}Q_{\,L)}^{M}\mathbf{x}_{M}\mathbf{x}_{J}+Q^{IJ}Q_{KL}\mathbf{x}_{I}\mathbf{x}_{J}}{\left|\mathbf{x}\right|^{d-1}}-\frac{2Q_{K}^{M}Q_{ML}}{\left(d-3\right)\left|\mathbf{x}\right|^{d-3}}\Bigg)\)} \\
& & \raisebox{3\height}{\(\Delta U^{\left(2\right)}=-\frac{8}{d^{2}-1}C_{2}^{E}Q_{IJ}Q^{IJ}\)} 

\end{tabular}
\caption{Feynman diagrams that contribute to the interaction energy in the gravito-electric background. The double line stands for the black hole worldline, the filled box is the interaction vertex with the fields, and crosses represent insertions of the background. The second column consists of the field correction and energy contribution of each diagram prior to renormalization according to the boundary conditions.}
\label{tab:gravito_electric_diagrams}
\end{table}

\subsection{Thermal corrections}
To determine the black-hole polarization induced by the gravito-electric background, we first evaluate the free and internal energies associated with the black-hole response. Combining \eqref{eq:action_perturbation_def}, \eqref{eq:free_energy_is_action}, and \eqref{eq:GSfullS} yields
\begin{align}
\label{eq:GE_free_energy}
F^{\left(2\right)}=-C_{2}^{E}Q^{2}.
\end{align}

To obtain the internal energy, we substitute the metric corrections from Table~\ref{tab:gravito_electric_diagrams} into \eqref{eq:U_with_NRG}. The contribution from each diagram appears in the second row of the last column of Table~\ref{tab:gravito_electric_diagrams}. The total interaction energy is obtained by summing these contributions and adding the background energy associated with \eqref{eq:gravito_electric_background_def}, as evaluated in \eqref{eq:U_GE_background}.

We next implement the boundary condition \eqref{eq:GS_bndry_cond}. As before, this amounts to an appropriate rescaling of $Q_{IJ}$. The details are presented in Appendix~\ref{app:full_action_GE_consistency}, where we find that the background energy \eqref{eq:U_GE_background} acquires a finite correction under this rescaling,
\begin{align}
\label{eq:U_QQ_EFT}
\overline{U}^{\left(2\right)}\rightarrow\overline{U}^{\left(2\right)}-\frac{4\left(d-7\right)}{\left(d-3\right)\left(d+1\right)}C_{2}^{E}Q^{IJ}Q_{IJ}.
\end{align}
Combining the energy values in Table~\ref{tab:gravito_electric_diagrams} with \eqref{eq:U_QQ_EFT}, we obtain the black hole gravito-electric interaction energy,
\begin{align}
\label{eq:U_total_QQ}
U^{\left(2\right)}=-\frac{\left(d-7\right)}{\left(d+1\right)}C_{2}^{E}Q^2-\frac{4\left(d-7\right)}{\left(d-3\right)\left(d+1\right)}C_{2}^{E}Q^2=-\frac{\left(d-7\right)}{\left(d-3\right)}C_{2}^{E}Q^2.
\end{align}
For \(d=4\) this reduces to \(U^{\left(2\right)}=3C_{2}^{E}Q^2\).

Finally, we determine the gravito-electric polarization of the black hole at quadratic order in the background.
Substituting \eqref{eq:U_total_QQ} and \eqref{eq:GE_free_energy} in \eqref{eq:QP_F_U} for the \(l=2\) case and identifying \(-Q_{IJ}\) as the control parameter, we obtain
\begin{align}
\label{eq:gravito_electric_TS}
\delta^{\left(2\right)}\left(TS\right)&=\frac{4}{d-3} C_{2}^{E}Q^2, \\ 
\label{eq:gravito_electric_QP}
\mathcal{P}^{IJ}&=\frac{\left(d-3\right)^2+4}{2\left(d-3\right)}C_{2}^{E}Q^{IJ}.
\end{align}
Where for \(d=4\) this amounts to \(\delta^{\left(2\right)}\left(TS\right)=4C_{2}^{E}Q^2\) and \(\mathcal{P}^{IJ}Q_{IJ}=\frac{5}{2}C_{2}^{E}Q^2=\frac{5}{8}\delta^{\left(2\right)}\left(TS\right)\).

\section{Discussion}
\label{sec:conclusions}

In this work, we perturbatively extended the thermodynamic formulation of black hole mechanics \cite{Bardeen:1973gs} to describe a Schwarzschild black hole in an arbitrary number of spacetime dimensions immersed in weak external background fields of various types. We also introduced a novel, gauge-invariant method for deriving the corresponding Love numbers.

To justify the perturbative treatment, we assume a hierarchy of scales in which the characteristic curvature radius of the external background is much larger than the black hole horizon size. This separation enables a systematic and controlled EFT description of the black hole response, parameterized by an infinite set of coefficients \cite{Goldberger:2004jt,Porto:2005ac,Goldberger:2007hy,Kol:2007rx,Kol:2007bc,Rothstein:2014sra,Porto:2016pyg,Levi:2018nxp,Goldberger:2022rqf} – the Love numbers.

In our approach, the Love numbers are obtained by evaluating the full gravitational action on classical solutions that asymptotically match a prescribed external background. This construction is manifestly gauge invariant, as it relies on the evaluation of the action itself – a gauge-invariant quantity – and avoids the need for gauge fixing or the computation of Feynman diagrams on the EFT side. From a classical EFT perspective, this procedure can be understood as integrating out short-wavelength modes in the presence of fixed long-wavelength modes, specified through asymptotic boundary conditions.

We first considered perturbations of the black hole spacetime induced by external matter fields, including scalar and Maxwell backgrounds, and then turned to pure gravitational perturbations of gravito-electric (scalar) and gravito-magnetic (vector) type. By evaluating the full gravitational action on shell as a functional of the asymptotic boundary conditions, we reproduced the Love numbers reported in the literature and derived thermal corrections to the free energy of the Schwarzschild black hole.

In addition, we introduced a thermodynamic polarization variable conjugate to the gauge-invariant control parameter associated with the external background field, and exploited the scale invariance of the equations of motion to derive a generalized Smarr relation \eqref{eq:TS_U_QP} in the presence of external perturbations.

Our approach extends the standard thermodynamic relations within a perturbative framework. A fully non-perturbative treatment, by contrast, may be obstructed by ambiguities that would need to be resolved. We have not pursued this direction here, leaving a non-perturbative definition of black hole polarization as an interesting avenue for future work.

Using the gravitational pseudotensor together with a diagrammatic EFT approach, we determined corrections to the internal energy of the unperturbed Schwarzschild black hole. Combining this result with the free energy and the generalized Smarr relation enabled the derivation of the black hole’s polarization tensor.

The expressions for the free and internal energies contain contributions that depend explicitly on the scale set by the geometric size of the boundary surface. These terms are sensitive to boundary rescalings, and most vanish as the boundary is taken to infinity, leaving only a small subset that remains unsuppressed. The surviving contributions correspond to the self-energy of the background and the interaction energy between the unperturbed Schwarzschild black hole and the background. Since these terms do not encode information about the induced polarization, they are subtracted, yielding a gauge-invariant quantity that is independent of boundary-related scales.

Our separation of the black-hole response from contributions determined solely by the prescribed external background and the unperturbed Schwarzschild geometry has a close conceptual parallel with the subtracted geometries introduced in \cite{Cvetic:2011hp,Cvetic:2011dn}. In those constructions, modifications of the metric alter the asymptotic structure of the geometry while preserving the horizon data and thermodynamic properties of the black hole. Remarkably, the resulting effective geometries exhibit an enhanced $SL(2,\mathbb R)$ structure, which is closely related to the Love symmetry underlying black-hole tidal responses \cite{Porto:2016zng,Charalambous:2021kcz,Hui:2021vcv,Hui:2022vbh,Charalambous:2022rre,Rai:2024lho,Combaluzier-Szteinsznaider:2024sgb,Kehagias:2024rtz,Gounis:2024hcm,Rodriguez:2026iot,Chakraborty:2026qru}. It would be interesting to formulate the thermodynamic polarization variables derived here in terms of this hidden symmetry and to investigate whether the subtracted-geometry description can provide a geometric interpretation of the response coefficients and associated polarization. The approach presented here, however, differs in an important respect: we do not modify the physical geometry, but rather isolate the response of the Schwarzschild black hole to external fields directly at the level of the on-shell action, by separating contributions associated with the black hole, the external background, and their interaction.

To leading order, the polarization tensor is proportional to an appropriate Love number, which vanishes identically in four dimensions but can take either sign in higher dimensions. Consequently, the corresponding correction to the entropy is not sign-definite, contrary to the naive expectation that an external field increases order in the system and thus reduces its entropy. However, this alternating sign does not necessarily indicate a thermodynamic instability, as Schwarzschild black hole thermodynamics is already unconventional even in the absence of perturbations – for example, it exhibits a negative heat capacity. It would be interesting to further investigate how this sign ambiguity affects the evaporation process.

Moreover, in certain dimensions, the Love numbers and the polarization tensor exhibit divergences. For instance, all $l=2$ Love numbers considered in this work diverge at $d=7$. In these special dimensions, the contributions of the mass and the Love numbers to various physical observables mix, giving rise to additional (formally divergent) contributions to the thermodynamics that must be properly accounted for.

These divergences cancel once all relevant terms are combined, yielding finite expressions with logarithmic dependence. Such logarithms signal classical renormalization group running of Love numbers \cite{Kol:2011vg,Hui:2020xxx,Hadad:2024lsf,Ivanov:2022hlo,Mandal:2023hqa,Barbosa:2025uau,Noumi:2026shc,Barbosa:2026qcv}, and a more detailed analysis of this structure would be an interesting direction for future work.

Furthermore, the computed corrections to the internal energy and entropy suggest a possible avenue for testing general relativity. As an illustration, consider the merger of two Schwarzschild black holes into a single remnant. The laws of black-hole mechanics imply a monotonically increasing entropy that scales as mass squared, while energy conservation is linearly proportional to the mass. Together, they impose both lower and upper bounds on the remnant mass
\begin{align}
\label{eq:entropy_merger_rises_and_energy}
\sqrt{M_1^2+M_2^2}\leq M_{1+2}\leq M_1+M_2,
\end{align}
where \(M_1\) and \(M_2\) are the initial black-hole masses and \(M_{1+2}\) is the mass of the final remnant. In a stationary background, apparent violations of these bounds may arise from two sources. The first are dynamical changes and radiation during the merger, which may alter both the energy and the entropy. The second source arises from internal black-hole structure, parametrized to leading order in this paper through polarization effects. The relations \eqref{eq:TS_U_QP} and \eqref{eq:QP_F_U} and their realizations for each type of background provide a framework for isolating black hole internal structure contributions. A crucial next step toward fully constraining the entropy bound in this setting is the computation of background-induced corrections to the temperature.

Finally, the vanishing of a black hole’s response to stationary perturbations in four spacetime dimensions does not necessarily survive possible corrections to Einstein gravity. The underlying Love symmetry may be broken by various modifications, such as higher-curvature corrections \cite{Cardoso:2018EFT,Wang:2026qst} or a cosmological constant \cite{Nair:2024dS,Franzin:2024cah}, as well as by semiclassical quantum gravity effects \cite{Brustein:2020tpg,Kim:2020Love,Brustein:2021bnw}. Our framework provides a possible avenue for exploring these scenarios. Certain thermodynamic concepts may require generalization to incorporate corrections to Einstein gravity – for example, the entropy functional \cite{Wald:1993nt}; however, the overall structure remains intact. Furthermore, introducing a finite cutoff offers a self-consistent way to address ambiguities in non-asymptotically flat spacetimes like de Sitter \cite{Nair:2024dS}. By evaluating the response on a boundary, the local tidal dynamics are effectively decoupled from the global geometry. This procedure avoids the need for asymptotic matching, which becomes ill-defined in the presence of a cosmological horizon, a complication relevant for all $d\ge4$.

\acknowledgments

We thank Walter Goldberger, Tomer Hadad, Barak Kol, and Ira Rothstein for insightful discussion and correspondence. We acknowledge partial support from the BSF Grant No. 2022113, NSF-BSF Grant No. 2022726 and ISF Grant No. 2526/25. Additionally, we acknowledge partial support from Israel’s Council for Higher Education.

\appendix
\section{Small variations}
\label{app:curvature_variations}

In this appendix, we gather all the necessary definitions and formulas to carry out the variations of the gravitational action that we calculate in the main body of the text. To begin with, we present the final result for the variation of the Ricci scalar under a metric perturbation, $g_{\mu\nu}\to g_{\mu\nu} + \delta g_{\mu\nu}$,
\begin{equation}
 \delta R = - R^{\mu\nu} \delta g_{\mu\nu} + 
 \nabla^\mu\big(\nabla^{\nu}\delta g_{\mu\nu} - g^{\alpha\beta}\nabla_{\mu}\delta g_{\alpha\beta} \big) + \mathcal{O}(\delta g^2)~.
\end{equation}

Next, consider the extrinsic curvature of a timelike hypersurface characterized by the embedding function $X^\mu(\sigma)$. It can be expressed as follows
\begin{equation}
    \mathcal{K}_{ac}=e^\mu_a e^\nu_c \nabla_\mu n_\nu ~, \quad n_\mu n^\mu= - 1~.
\end{equation}
where $n^\mu$ is a spacelike unit vector normal to the surface, and 
\begin{equation}
    e^\mu_a={\partial X^\mu\over \partial \sigma^a} ~, \quad \forall \sigma^a: \quad n_\mu e^\mu_a =0 ~.
\end{equation}
Using the orthogonality condition, we can rewrite $\mathcal{K}_{ac}$ as follows
\begin{equation}
    \mathcal{K}_{ac}=-  e^\mu_a n_\nu \nabla_\mu e^\nu_c 
 = -  n_\nu \Big({\partial^2 X^\nu \over \partial\sigma^a\partial\sigma^c} + \Gamma_{\mu\rho}^\nu {\partial X^\mu \over \partial\sigma^a}
 {\partial X^\rho \over \partial\sigma^c}\Big) ~.
\end{equation}
Under a metric perturbation, we have 
\begin{equation}
    \quad n_\nu \to n_\nu + \delta n_\mu~, 
    \quad e^\mu_a \to e^\mu_a~.
\end{equation}
These are exact relations. To preserve the orthogonality condition, we must have
\begin{equation}
    0 =  e^\mu_a (n_\mu + \delta n_\mu) = e^\mu_a \delta n_\mu ~.
\end{equation}
Hence, the normal vector exhibits changes along $n^\mu$ only.
Note that this holds to all orders in $\delta g_{\mu\nu}$. To express $\delta n_\mu$ in terms of $\delta g_{\mu\nu}$, we impose the normalization condition 
\begin{equation}
    -1=(g^{\mu\nu}-\delta g^{\mu\nu} + \ldots) (n_\mu+\delta n_\mu) (n_\nu+\delta n_\nu) ~,
\end{equation}
where indices are raised and lowered with the unperturbed metric, {\it e.g.}, $\delta g^{\mu\nu} = g^{\mu\alpha} g^{\nu \beta} \delta g_{\alpha \beta}$, and ellipsis encode quadratic and higher order terms in $\delta g_{\mu\nu}$. Thus,
\begin{equation}
    n^\mu \delta n_\mu = {1\over 2} \delta g^{\mu\nu} n_\mu n_\nu + \mathcal{O}(\delta g^2) ~.
\end{equation}
Combining, yields
\begin{equation}
     \delta n_\mu = -{1\over 2}\, n_\mu  \, n^\alpha n^\beta \delta g_{\alpha\beta}   + \mathcal{O}(\delta g^2) ~.
\end{equation}
Varying the above expression one more time gives 
\begin{equation}
     \delta^\mt{(2)} n_\mu = n_\mu\Big( g^{\mu\nu} n^\alpha n^\beta \delta g_{\mu \alpha} \delta g_{\nu\beta}
     + {3\over 4} \big(n^\alpha n^\beta \delta g_{\alpha\beta}\big)^2 \Big)  ~.
\end{equation}
Hence,
\begin{equation}
    \delta n_\mu = {1\over 2}n_\mu \Big(  - n^\alpha n^\beta \, \delta g_{\alpha\beta}
     + g^{\mu\nu} n^\alpha n^\beta \delta g_{\mu \alpha} \delta g_{\nu\beta}
     + {3\over 4} \big(n^\alpha n^\beta \delta g_{\alpha\beta}\big)^2 \Big) 
    + \mathcal{O}(\delta g^3) ~.
\end{equation}
The induced metric in terms of the ambient metric is given by
\begin{equation}
    h_{ac}= e^\mu_a e^\nu_c g_{\mu\nu} \quad \Rightarrow \quad
    \delta h_{ac} = e^\mu_a e^\nu_c \delta g_{\mu\nu} ~,
\end{equation}
and the following expansions hold under small variations of the metric 
\begin{align}
    \sqrt{h}&\to \sqrt{h}\Big(1+{1\over 2} h^{ac}\delta h_{ac} + {1\over 8} (h^{ac}\delta h_{ac})^2 
    - {1\over 4}\, \delta h^{ac} \, \delta h_{ac} 
    + \mathcal{O}(\delta h^3)\Big) ~,
    \nonumber \\
h^{ac} &\to h^{ac} - \delta h^{ac} + \delta h^{a}_b \delta h^{bc} + \mathcal{O}(\delta h^3) ~,
\end{align}
where indices of $\delta h_{ac}$ are raised and lowered with the unperturbed induced metric.

It is convenient to employ $e_a^\mu$ to transform $h^{ab}$ and $\mathcal{K}^{ab}$ into bulk tensors,
\begin{equation}
 h^{\mu\nu}=h^{ac} e^\mu_a e^\nu_c ~, \quad 
 \mathcal{K}^{\mu\nu}=\mathcal{K}^{ac} e^\mu_a e^\nu_c = 
 h^{\mu\alpha}h^{\nu\beta}\nabla_\alpha n_\beta ~.
\end{equation}
Note also that the following relations hold
\begin{align}
 h^{\mu\nu}&=g^{\mu\nu} + n^\mu n^\nu ~,
 \nonumber \\
 0&= \nabla_\mu(n^\nu n_\nu)=2n^\nu\nabla_\mu( n_\nu) 
 \quad \Rightarrow \quad
 \mathcal{K}=h_{\mu\nu}\mathcal{K}^{\mu\nu}=g_{\mu\nu}\mathcal{K}^{\mu\nu} ~.
\end{align}
Using these definitions, it is straightforward to vary the extrinsic curvature
\begin{align}
 \delta \mathcal{K}_{\mu\nu} =& - {1\over 2}\, n^\alpha n^\beta \delta g_{\alpha\beta} \mathcal{K}_{\mu\nu}
 -\delta g_{\alpha\beta} n^\beta \big(n_\mu \mathcal{K}^\alpha_\nu + n_\nu \mathcal{K}^\alpha_\mu \big)
 \nonumber \\
 &-{1\over 2} h_\mu^\beta h_\nu^\rho n^\alpha \big(\nabla_\beta \delta g_{\alpha \rho}+\nabla_\rho \delta g_{\alpha \beta} - \nabla_\alpha \delta g_{\beta \rho}\big)
  + \mathcal{O}(\delta g^3) ~.
  \label{eq:K}
\end{align}
Similarly, 
\begin{equation}
 \delta \mathcal{K} = -{1\over 2}\mathcal{K}^{\mu\nu}\delta g_{\mu\nu} 
 - {1\over 2}n^\mu\big(\nabla^\nu\delta g_{\mu\nu} - g^{\alpha\beta}\nabla_\mu \delta g_{\alpha\beta}\big) 
  - \, {1\over 2}h_{\mu\nu} \nabla^\mu V^\nu + \mathcal{O}(\delta g^2)~,
  \label{eq:traceK}
\end{equation}
where
\begin{equation}
 V^\nu= n^\nu n^\alpha n^\beta \delta g_{\alpha\beta} +  g^{\nu\alpha} n^\beta \delta g_{\alpha\beta} ~. 
\end{equation}
Note that, by definition, $V^\mu$ is tangent to the timelike hypersurface formed by the spacelike unit vector $n^\mu$. Therefore, the last term in \eqref{eq:traceK} represents a covariant divergence of $V^\mu$ restricted to the hypersurface, and its integral over this surface vanishes.

\section{Useful integrals}
\label{apx:integrals}

Using integration by parts, the tracelessness of tensors associated with background perturbations, and the simple relation $\partial_I^x|x-y|=-\partial_I^y|x-y|$, the evaluation of all diagrams in this work can be reduced to derivatives of a single master integral of the form 
\begin{equation}
 \int d^{D} x \frac{1}{ |x|^{\alpha} |x-y|^\beta}
 =\pi^{D/2} {\Gamma\big({\alpha+\beta-D \over 2}\big)~\Gamma\big({D-\alpha\over 2}\big)~\Gamma\big({D-\beta\over 2}\big) \over \Gamma\big({\alpha\over 2}\big)~\Gamma\big({\beta\over 2}\big)~\Gamma\big(D-{\alpha+\beta\over 2}\big)} ~|y|^{D-\alpha-\beta} ~.
 \label{appx:I}
\end{equation}

In addition, we require the spherical integral
\begin{equation}
   \int d\Omega ~ x^{J_{1}}...x^{J_{l}} x^{I_{1}}...x^{I_{l}} 
   = N \, r^{2l} \, \big(\delta^{I_1J_1} \cdots\delta^{I_lJ_l} + \ldots\big)  ~,
   \label{use_iden1}
\end{equation}
where the right-hand side is fixed by rotational symmetry, and $N$ is a dimensionless constant to be determined. The ellipsis denotes the distinct contractions formed from products of $l$ Kronecker delta functions, arranged such that the resulting tensor is completely symmetric in the indices $I_1, \ldots, I_l, J_1,\ldots, J_l$. The total number of distinct products is $(2l)!/(l! 2^l)$. In particular, choosing all indices to be identical, say 1, yields 
\begin{equation}
    \int d\Omega ~ (x^1)^{2l} = N \, r^{2l} {(2l)! \over l! 2^l}~.
\end{equation}
Or equivalently,
\begin{equation}
    \Omega_{d-2}\int_0^\pi d\theta \sin^{d-3}\theta \cos^{2l}\theta = N  {(2l)! \over l! 2^l} ~.
\end{equation}
Substituting
\begin{equation}
 \int_0^\pi d\theta \sin^{d-3}\theta \cos^{2l}\theta = {\Gamma\big({d-2\over 2}\big) \Gamma\big({2l+1\over 2}\big) \over \Gamma\big({d-1+2l\over 2}\big)} ~,
\end{equation}
gives
\begin{equation}
 N={\pi^{d-1\over 2} \over 2^{l-1} \Gamma\big({d-1+2l\over 2}\big)} = {\Omega_{d-1+2l}\over (2\pi)^l} ~.
 \label{use_iden2}
\end{equation}

\section{Background energy}
\label{appx:bckgrnd_energy}

If the black hole in our setup is removed – for instance, by setting $m = 0$ – the resulting spacetime is not flat due to the presence of a nontrivial weakly curved background. In this appendix, we evaluate its energy to leading order in the perturbation. Since the linear-order contribution vanishes in all cases considered, we focus on the quadratic correction.

In the case of a minimally coupled scalar field governed by the actions \eqref{eq:EH_action} and \eqref{eq:min_scalar_action}, the Einstein equations of motion in the static limit ($A_I = 0$) take the form
\begin{equation}
    \mathcal{R}[\gamma]=-{2(d-2)\over (d-3)^2}\gamma^{IJ}T^\phi_{IJ}+
    8\pi \gamma^{IJ}\partial_I\psi\partial_J\psi ~, \quad \partial_I \big( \sqrt{\gamma} \, \gamma^{IJ} \,\partial_J\phi\big)=0~.
\end{equation}
These equations are exact. However, we are interested in a weakly curved background induced by \eqref{eq:scalar_bckrnd}. In this setup, $\overline{\phi} = 0$ to all orders in $\Psi_{J_1 \cdots J_l}$, while the leading nontrivial correction to the background metric, $\overline{\sigma}_{IJ}$, arises at quadratic order and satisfies\footnote{Throughout this appendix, indices on tensors involving metric perturbations are raised and lowered with the Kronecker delta $\delta_{IJ}$.}
\begin{equation}
    \partial^I \partial^J\overline{\sigma}^{(2)}_{IJ}-\partial^2\overline{\sigma}^{(2)}=8\pi(\partial\overline{\psi})^2 ~.
\end{equation}
Substituting this result into \eqref{eq:U_with_NRG}, we obtain the background energy at quadratic order in the perturbation \eqref{eq:scalar_bckrnd}
\begin{align}
    \overline{U}^{(2)}&={1\over 16\pi}\int \big(\partial_J \overline{\sigma}_{I}^{(2)J} -\partial_I\overline{\sigma}^{(2)}\big) n^I R^{d-2}d\Omega 
  = {1\over 16\pi}\int d^{d-1}x\big(\partial^I \partial^J\overline{\sigma}^{(2)}_{IJ}-\partial^2\overline{\sigma}^{(2)}\big)
  \nonumber \\
  &={1\over 2}\int d^{d-1}x (\partial\overline{\psi})^2 = {\Omega_{d-1}\over 2(d-1)}R^{d-1}\Psi_J\Psi^J~,
\end{align}
where in the final step the integral has been evaluated for the case $l = 1$.

Similarly, for electromagnetic matter, the total action is given by the sum of \eqref{eq:EH_action} and \eqref{eq:Maxwell_action}. In the static limit ($A_I = \mathcal{A}_I = 0$), the exact Einstein equations of motion take the form
\begin{align}
    &\mathcal{R}[\gamma]=-{2(d-2)\over (d-3)^2}\gamma^{IJ}T^\phi_{IJ}- 
    8\pi e^{-2\phi}\gamma^{IJ}\partial_I\mathcal{A}_0\partial_J\mathcal{A}_0 ~, 
    \nonumber \\
    &{1\over\sqrt{\gamma}}\partial_I \big( \sqrt{\gamma} \, \gamma^{IJ} \,\partial_J\phi\big)= 8\pi {d-3\over d-2}\gamma^{IJ}\partial_I\mathcal{A}_0\partial_J\mathcal{A}_0 ~.
\end{align}
The weakly curved background induced by \eqref{eq:Maxwell_bckrnd} first appears at quadratic order in the perturbation, where the metric components satisfy
\begin{equation}
    \partial^I \partial^J\overline{\sigma}^{(2)}_{IJ}-\partial^2\overline{\sigma}^{(2)}=-8\pi\big(\partial\overline{\mathcal{A}}_0\big)^2 ~, \quad \partial^2\overline{\phi}^{(2)}=8\pi {d-3\over d-2} \big(\partial\overline{\mathcal{A}}_0\big)^2 ~.
\end{equation}
Substituting these expressions into \eqref{eq:U_with_NRG}, we obtain the background energy at quadratic order in the perturbation \eqref{eq:Maxwell_bckrnd}
\begin{align}
    \overline{U}^{(2)}&={1\over 16\pi}\int \Big(2{d-2\over d-3}\partial_I\overline{\phi}^{(2)}+\partial_J \overline{\sigma}_{I}^{(2)J} -\partial_I\overline{\sigma}^{(2)}\Big) n^I R^{d-2}d\Omega 
    \nonumber \\
  &= {1\over 16\pi}\int d^{d-1}x\Big(2{d-2\over d-3}\partial^2\overline{\phi}^{(2)}+\partial^I \partial^J\overline{\sigma}^{(2)}_{IJ}-\partial^2\overline{\sigma}^{(2)}\Big)
  \nonumber \\
  &={1\over 2}\int d^{d-1}x \big(\partial\overline{\mathcal{A}}_0\big)^2 = {\Omega_{d-1}\over 2(d-1)}R^{d-1}\mathcal{E}_{J}\mathcal{E}^{J}~,
\end{align}
where, as before, the final expression is obtained by evaluating the integral for the case $l = 1$.

Now consider the weakly curved gravito-electric background \eqref{eq:gravito_electric_background_def}. While the linear-order correction to the energy vanishes, a nontrivial contribution arises at second order, which we now proceed to evaluate. As before, staticity implies that $A_I = 0$ to all orders.

The spatial part of the metric receives corrections at second order in the perturbation, while the first correction to $\bar{\phi}$ appears only at cubic order. Since we compute the energy up to second order, this contribution can be neglected. The equations of motion \eqref{eq:EOM_NRG} can then be recast as
\begin{equation}
    \mathcal{R}[\gamma]={d-2\over d-3} \gamma^{IJ}\partial_I\phi\partial_J\phi \quad \Rightarrow \quad 
    \partial^I\partial^J\overline\sigma_{IJ}^{(2)}-\partial^2\overline\sigma={d-2\over d-3} (\partial\overline\phi)^2 ~.
\end{equation}
Substituting into \eqref{eq:U_with_NRG}, we obtain
\begin{equation}
    \label{eq:U_GE_background}
    \overline{U}^{(2)}={1\over 16\pi} {(d-2)(d-7)\over (d-3)^2} \int \big(\overline\phi \partial_I \overline\phi\big) n^I R^{d-2}d\Omega
    ={\Omega_{d-1}\over 4\pi} {(d-2)(d-7)\over (d-3)^2(d-1)(d+1)} R^{d+1}Q_{IJ}Q^{IJ} ~,
\end{equation}
where the integral over the sphere has been evaluated for the case $l = 2$ using \eqref{use_iden1}.

Finally, in the case of the gravito-magnetic perturbation \eqref{eq:gravito-magnetic-bckgnd}, one must expand the exact equations of motion \eqref{eq:EOM_NRG} to quadratic order. This yields
\begin{equation}
    \partial^I \partial^J\overline{\sigma}^{(2)}_{IJ}-\partial^2\overline{\sigma}^{(2)}=-{d-5\over d-3}\overline{F}^2 ~, \quad \partial^2\overline{\phi}^{(2)} =- \overline{F}^2 ~.
\end{equation}
Substituting these expressions into \eqref{eq:U_with_NRG} and rearranging terms, we obtain
\begin{equation}
    \overline{U}^{(2)}={1\over 16\pi}\int \Big(\partial_J\overline{A}_I+\partial_I \overline{A}_J \Big) \overline{A}^Jn^I R^{d-2}d\Omega 
  = {\Omega_{d-1}\over 4\pi(d-1)(d+1)}R^{d+1}C_{IJK}C^{IJK}~,
\end{equation}
where, in the final step, the integral has been evaluated for the case $l = 2$ using \eqref{use_iden1}.

\section{Consistency check}
\label{app:full_action_GE_consistency}
In this appendix, we perturbatively evaluate the full action \eqref{eq:full_exact_action} on shell for gravito-electric perturbations \eqref{eq:gravito_electric_background_def} of the Schwarzschild black hole, using the metric obtained within the EFT framework. The resulting expression agrees with the electric sector of \eqref{eq:PP_action}, providing a consistency check of the relation \eqref{eq:action_perturbation_def}.

%We start with the relation \eqref{eq:free_energy_is_action} together with the perturbative expansion of the full action given in \eqref{eq:action_perturbation_def}, \eqref{eq:d1S_full}, and \eqref{eq:d2S_full}. We impose the gravito-electric background \eqref{eq:gravito_electric_background_def} both as a boundary condition at \(r=R\) and as a perturbation of the Schwarzschild geometry.

The first-order corrections to the Schwarzschild metric induced by the gravito-electric perturbation \eqref{eq:gravito_electric_background_def} take the form
\begin{eqnarray}
\label{eq:g1components_GE}
&&g_{tt}^{\left(1\right)}=2f\phi, \,\,\, g_{rr}^{\left(1\right)}=\frac{2}{\left(d-3\right)}f^{-1}\phi -f^{-\frac{1}{d-3}} \sigma_{rr}, \non
&& g_{ij}^{\left(1\right)}=\frac{2}{\left(d-3\right)}r^{2}\Omega_{ij}\phi - f^{\frac{-1}{d-3}}\sigma_{ij} ~,
 g_{ir}^{(1)}=-f^{\frac{-1}{d-3}}\sigma_{ir} ~.
\end{eqnarray}
We note that terms in \(\delta^{(2)}S_{\textrm{full}}\) that depend only on fields evaluated on the boundary or on their tangential derivatives encode fixed boundary data and do not probe the black hole's response. Such contributions are therefore irrelevant for the interaction energy. Consequently, only boundary terms involving radial derivatives of the fields contribute, implying that the relevant part of the second variation \eqref{eq:d2S_full} arises solely from terms containing \(\delta \mathcal{K}\) and \(\delta \mathcal{K}_{\mu\nu}\),
\begin{align}
\label{eq:d2S_relevant_GE}
\delta^{\left(2\right)}S=
\frac{1}{16\pi}\int_{\partial V}\sqrt{\left|h\right|}\left(h^{\mu\nu}\delta^{(1)} \mathcal{K}-g^{\mu\alpha}g^{\beta\nu}\delta^{(1)} \mathcal{K}_{\alpha\beta}\right) g^{(1)}_{\mu\nu}
+\ldots~.
\end{align}
Here and throughout the remainder of this appendix, the ellipses denote irrelevant second-order terms. Using \eqref{eq:K} and \eqref{eq:traceK}, we similarly conclude that
\begin{align}
\label{eq:dK_relevant_GE}
\delta^{(1)} \mathcal{K}_{\mu\nu}&=\frac{1}{2}h_{\mu}^{\beta}h_{\nu}^{\rho}n^{r}\nabla_{r} g^{(1)}_{\beta\rho} + \ldots, \\
\delta^{(1)} \mathcal{K}&=\frac{1}{2}n^{r}\Big( g^{tt}\nabla_{r} g^{(1)}_{tt}+g^{ij}\nabla_{r} g^{(1)}_{ij}\Big) + \ldots ~.
\end{align}

Since the Christoffel symbols correspond to the unperturbed Schwarzschild metric, only partial radial derivatives in the above expression encode the black hole's response to the perturbation. Furthermore, since the Schwarzschild metric is diagonal, the expression \eqref{eq:d2S_relevant_GE} simplifies considerably, yielding
\begin{align}
\label{eq:d2S_relevant_GE_simple}
\delta^{\left(2\right)}S&=-\frac{1}{8\pi}\frac{\left(d-2\right)}{\left(d-3\right)}fR^{d-2}\int_{\partial V}d\Omega_{d-2}\phi\partial_{r}\phi \non &-\frac{1}{32\pi}f^{\frac{d-5}{d-3}}R^{d-6}\int_{\partial V}d\Omega_{d-2}\left(\Omega^{in}\Omega^{jm}-\Omega^{ij}\Omega^{mn}\right)\sigma_{mn}\partial_{r}\sigma_{ij}+\ldots ~.
\end{align}
The expression in the second line does not contribute an $R$-independent term to $\delta^{\left(2\right)}S$, since $\sigma_{ij}$ decays faster than $\phi$ at infinity. 

Up to this point, the boundary condition \eqref{eq:GS_bndry_cond} has not been explicitly imposed in the EFT calculation. To consistently use the metric corrections obtained in the EFT together with \eqref{eq:d2S_relevant_GE_simple}, we must ensure that \eqref{eq:GS_bndry_cond} is satisfied.
We therefore require the correction to the scalar field \(\phi\) to obey \eqref{eq:GS_bndry_cond}. 
This requirement fixes the relation between the amplitude 
\((Q_{IJ})_{\textrm{EFT}}\) used in the EFT and its counterpart measured at the boundary $r=R$. Implementing this condition amounts simply to a rescaling of the amplitude.
\begin{align}
\label{eq:Q_EFT}
Q^{IJ}=\left(Q^{IJ}\right)_{\textrm{EFT}}\left(1+C_{2}^{E}\frac{8\pi\left(d-3\right)\left(d-1\right)}{\left(d-2\right)\Omega_{d-1}R^{d+1}}
%+\left(Q_{IJ}\right)_{\textrm{EFT}}\frac{4\left(d+1\right)n^{I}n^{J}}{\left(d-2\right)\Omega_{d-3}R^{d-1}}
\right)
\end{align}
which implies
\begin{align}
\label{eq:QQ_EFT}
\left(Q^{IJ}\right)_{\textrm{EFT}}\left(Q_{IJ}\right)_{\textrm{EFT}}\simeq Q^{IJ}Q_{IJ}\left(1-C_{2}^{E}\frac{16\pi\left(d-3\right)\left(d-1\right)}{\left(d-2\right)\Omega_{d-1}R^{d+1}}\right)
\end{align}
Here the expansion has been truncated consistently, neglecting higher powers of \(Q\) terms and subleading terms in \(1/R\).

To isolate terms proportional to \(C_2^{E}Q^{2}\), one factor of \(\phi\) in \eqref{eq:d2S_relevant_GE_simple} is replaced by the background field \eqref{eq:gravito_electric_background_def}, while the other is taken from the EFT-generated correction listed in Table~\ref{tab:gravito_electric_diagrams}(a). Although inserting the background twice would naively yield no $R$-independent contribution, rescaling the background according to \eqref{eq:QQ_EFT} gives rise to an additional finite term. Collecting all such terms, we obtain
\begin{align}
\label{eq:S_relevant_GE_final}
S=\frac{1}{2}\delta^{\left(2\right)}S=C_{2}^{E}Q^{2}\left(\frac{\left(d-1\right)}{\left(d+1\right)}-\frac{2}{\left(d+1\right)}+\frac{4}{\left(d+1\right)}\right)=C_{2}^{E}Q^{2},
\end{align}
in agreement with the effective point-particle action \eqref{eq:PP_action_weak_fields} evaluated on the background \eqref{eq:gravito_electric_background_def} with \(l=2\). This provides a nontrivial consistency check of the approach in general and of the rescaling procedure in particular.

\bibliographystyle{utphys.bst}
\bibliography{Refs.bib}

\end{document}